\documentclass[prd,twocolumn,floatfix,amsmath,nofootinbib,amssymb,floatfix]{revtex4}
\usepackage{graphicx,color,dcolumn,booktabs,bm}
\usepackage{longtable,lscape}
\usepackage{pdfpages}
\usepackage{txfonts}
\usepackage{overpic}
\usepackage{amssymb}
\usepackage{makecell}
\usepackage{indentfirst}
\usepackage{feynmf}   
\usepackage{slashed}  
\usepackage{cases}
\usepackage{color}
\usepackage{multirow}
\usepackage{threeparttable}
\usepackage{epstopdf}
\usepackage{enumerate}
\usepackage{subfigure}
\usepackage{textcomp}
\usepackage{diagbox}
\usepackage{graphicx,color,dcolumn,booktabs,bm}
\usepackage{mathrsfs}
\usepackage{cancel}
\usepackage{float}
\usepackage[misc]{ifsym}
\usepackage[colorlinks,
citecolor=blue,
anchorcolor=red,
menucolor=red,
linkcolor=red,
filecolor=red,
runcolor=red,
urlcolor=blue,
frenchlinks=red]{hyperref}
\usepackage{array}
\usepackage{booktabs}

\begin{document}
\title{The form factors of transition process $\mathcal{B}_{Q_{1}Q_{2}}(\frac{1}{2}^{+})\rightarrow\mathcal{B}_{Q_{1}}^{*}(\frac{3}{2}^{+})$ and their applications to semileptonic decays}
\author{Guo-Liang Yu$^{1,3}$}
\email{yuguoliang2011@163.com}
\author{Peng Yang$^{1}$}
\author{Zhi-Gang Wang$^{1,3}$}
\email{zgwang@aliyun.com}
\author{Jie Lu$^{2}$}
\email{l17693567997@163.com}
\author{Bin Wu$^{2}$}
\author{Ze Zhou$^{1}$}

\affiliation{$^1$ Department of Mathematics and Physics, North China
Electric Power University, Baoding 071003, People's Republic of
China\\$^2$ School of Physics, Southeast University, Nanjing 210094, People's Republic of
China\\$^3$ Hebei Key Laboratory of Physics and Energy Technology,
North China Electric Power University, Baoding 071000, China}
\date{\today }

\begin{abstract}
In the framework of QCD sum rules, we obtain the form factors of the weak transition process $\mathcal{B}_{Q_{1}Q_{2}}(\frac{1}{2}^{+})\rightarrow\mathcal{B}_{Q_{1}}^{*}(\frac{3}{2}^{+})$. In this analysis, we eliminate the interferences of the low spin states and negative parity states to final results by employing 16 different dirac structures. As an application, these form factors are finally used to analyze the semileptonic decays of $\mathcal{B}_{Q_{1}Q_{2}}(\frac{1}{2}^{+})\rightarrow\mathcal{B}_{Q_{1}}^{*}(\frac{3}{2}^{+})l\nu$, where these decays are driven by the transition processes $c\rightarrow d/s+l^{+}+\nu_{l}$ and $b\rightarrow u+l^{-}+\overline{\nu}_{l}$. The predicted physical quantities include not only the partial widths, ratios of $\Gamma_{L}/\Gamma_{T}$ and the branching fractions, but also some observables such as the forward-backward asymmetry parameter $A_{FB}^{l}$ of lepton, the $P_z^{F}$ component of the polarization vector for daughter baryon and the longitudinal polarization of the lepton $P_z^{l}$. We hope all of these theoretical predictions about the weak decays will be helpful for studying the properties of doubly heavy baryons in experiments in the future.
\end{abstract}


\maketitle

\section{Introduction}\label{sec1}

The doubly heavy baryons which are composed of two heavy quarks and one light quark is an excellent laboratory to study the interaction between heavy and light quarks. In theory, physicists employed different approaches such as the relativistic or non-relativistic quark models \cite{Ebert:2002ig,Roberts:2007ni,Yu:2022lel,Li:2022ywz,Yu:2022ymb,Li:2022oth,Li:2022xtj,Karliner:2014gca,Valcarce:2008dr}, QCD sum rules (QCDSR)\cite{Bagan:1992za,Wang:2010hs,Wang:2010vn,Kiselev:2001fw,Zhang:2008rt,Aliev:2012ru}, bag model \cite{He:2004px,Cheng:2026mlv} and lattice QCD \cite{Lewis:2001iz,Flynn:2003vz,Liu:2009jc,Meinel:2021mdj} to analyze the mass spectra, the strong and weak decay properties of these baryons. These theoretical researches have provided important information to deepen our understanding about the inner structure of doubly heavy baryons and the heavy-quark symmetry.

In experiments, the first breakthrough is the discovery of the doubly charmed baryon $\Xi_{cc}^{++}$ by the LHCb collaboration \cite{352117} in 2017. This state was firstly observed in the decay channel $\Xi_{cc}^{++}\rightarrow \Lambda_{c}^{+}K^{-}\pi^{+}\pi^{+}$ with a measured mass $3621.40\pm0.72\pm0.14\pm0.27$ MeV and later was confirmed in another decay mode $\Xi_{cc}^{++}\rightarrow \Lambda_{c}^{+}\pi^{+}$ \cite{352118,352119}. Besides, experimental physicists also devoted themselves to searching for the bottom-charm and doubly bottom systems. For example, the LHCb collaboration tried to search for the doubly heavy baryon $\Xi_{bc}^{0}$ in the $D^{0}pK$ invariant mass spectrum \cite{LHCb:2020iko}, and
to search for the $\Omega_{bc}^{0}$ and $\Xi_{bc}^{0}$ in the $\Lambda^{+}_{c}\pi^{-}$ and $\Xi^{+}_{c}\pi^{-}$ mass spectra \cite{LHCb:2021xba}. However, only suggestive hints of signals have been detected, which implies the unobserving of these states.

Although the current experimental results in searching for bottom-charm and doubly bottom baryons are not so optimistic, we still believe that there will be a breakthrough in this field with the efforts of physicists. The firstly discovered heavy baryons are usually the ground states which can be reconstructed by the weak decay final states. Among these weak decays, the semi-leptonic decays are suitable for analyzing the QCD dynamics of doubly heavy baryons theoretically, since all the QCD dynamics can be
expressed by the hadron transition matrix elements. These matrix elements can be decomposed into several form factors. Thus, the accurate calculation of the form factor is very important for us to understand the structure and dynamics of the doubly heavy baryons. Besides, precise knowledge of the form factor is also essential for extracting the fundamental parameters such as the Cabibbo-Kobayashi-Maskawa (CKM) matrix elements which is a crucial step in searching for new physics beyond the Standard Model (SM).

Up to now, many semileptonic decay process of the heavy baryons have already been studied with the quark models \cite{Faustov:2018ahb,Geng:2020ofy,Albertus:2004wj,Ebert:2006rp,Cheng:1996cs,Ivanov:1997hi,Ivanov:1997ra,Faessler:2009xn,Gutsche:2018utw,Gutsche:2019iac}, the flavor symmetry method \cite{Wang:2017azm,Albertus:2013wja}, the light-front quark model \cite{Zhao:2018mrg,Zhao:2018zcb,Chua:2018lfa,Chua:2019yqh,Ke:2019smy,Zhu:2018jet,Li:2021qod,Hu:2020mxk,Li:2021kfb,Lu:2023rmq,Liu:2022mxv}, and QCDSR or light cone QCD sum rules (LCSR) \cite{Aliev:2010uy,Aliev:2023tpk,Azizi:2011mw,Wang:2008sm,Wang:2015ndk,Khodjamirian:2011jp,Zhao:2020mod,Shi:2019hbf,Khajouei:2024frw,Miao:2022bga,
Xing:2021enr,Zhao:2021sje,Lu:2026qkk,Zhang:2023nxl,Neishabouri:2024gbc,Tousi:2024usi}. Most of these above literatures concentrate on spin $\frac{1}{2}\rightarrow\frac{1}{2}$ transition process, few works focus on $\frac{1}{2}\rightarrow\frac{3}{2}$ process with the initial state being bottom-charm or doubly bottom baryons. In our previous work \cite{Yu:2026tbk}, we analyzed the form factors and semileptonic decays of $\Xi_{cc}$ and $\Omega_{cc}$ driven by the $c\rightarrow s/d$ transition at the quark model. As a continuation of this work, we will analyze the form factors and the semileptonic decays of $\Xi_{bb}$, $\Omega_{bb}$, $\Xi_{bc}$ and $\Omega_{bc}$ baryons, where the transition processes are driven by the $c\rightarrow s/d$ and $b\rightarrow u$. To provide valuable information for high-energy experiments, some obserbables will also be predicted such as the the forward-backward asymmetry parameter $A_{FB}^{l}$ of lepton, the $P_z^{F}$ component of the polarization vector for daughter baryons and the longitudinal polarization of the lepton $P_z^{l}$.

In the field of studying weak decay process, QCDSR is a very effective non-perturbative method which has been widely used to calculate the mass spectra, pole residues, strong coupling constants~\cite{Shifman:1978bx,Shifman:1978by,Colangelo:2000dp,Wang:2025sic,Wang:2022ufh,Wang:2018lhz,Wang:2017vtv,Zhang:2025qmg,Lu:2025zaf}, and the form factors \cite{Wang:2008sm,Wang:2015ndk,Khodjamirian:2011jp,Zhao:2020mod,Shi:2019hbf,Azizi:2011mw,Aliev:2023tpk,Khajouei:2024frw,Miao:2022bga,
Xing:2021enr,Zhao:2021sje,Zhang:2023nxl,Neishabouri:2024gbc,Tousi:2024usi}.
On the phenomenological side, the three-point correlation function for $\mathcal{B}_{Q_{1}Q_{2}}(\frac{1}{2}^{+})\rightarrow\mathcal{B}_{Q_{1}}^{*}(\frac{3}{2}^{+})$ can be decomposed into many dirac structures when the contributions of the low spin and negative parity states are considered. Under this condition, the form factors $F_{i}$ and $G_{i}$ ($i=1\sim4$) are not independent on each other and each form factor can not be determined only according to one dirac structure. Thus, we establish 16 linear equations to extract the form factors and to eliminate the contaminations of the low spin and negative parity states.

This article is organized as follows. After introduction in Sec. \ref{sec1}, we firstly introduce how the form factors of the transition process $\mathcal{B}_{Q_{1}Q_{2}}(\frac{1}{2}^{+})\rightarrow\mathcal{B}_{Q_{1}}^{*}(\frac{3}{2}^{+})$ are extracted in Sec. \ref{sec2}. In Sec. \ref{sec3}, we display the numerical results of the form factors and compare our results with those of other collaborations. Sec. \ref{sec4} is employed to discuss the semileptonic decay process, where the partial width, branching fractions and some observables are predicted. Sec. \ref{sec5} is the conclusion. Some complicated formulas and figures are shown in Appendices~\ref{Sec:AppA} and \ref{Sec:AppB}.

\section{The transition form factors of doubly bottom and bottom-charm baryons for $\frac{1}{2}\rightarrow\frac{3}{2}$ process}\label{sec2}
The semileptonic decay process $\mathcal{B}_{Q_{1}Q_{2}}(\frac{1}{2}^{+})\rightarrow\mathcal{B}_{Q_{1}}^{*}(\frac{3}{2}^{+})l\nu$ is related to the following transition matrix element,
\begin{eqnarray}\label{eq:3}
\notag
&&\left\langle \mathcal{B}_{Q_{1}}^{*}\left(p^{\prime}\right)|\overline{q}^{\prime}\gamma_{\nu}(1-\gamma_{5})Q_{2}|\mathcal{B}_{Q_{1}Q_{2}}^{}\left(p\right) \right\rangle \\ \notag
&&=\overline{u}_{\alpha}^{\mathcal{B}_{Q_{1}}^{*}}\left(p^{\prime},s^{\prime}\right)\left[\gamma_{\nu}\frac{p_{\alpha}}{m_{\mathcal{B}_{Q_{1}Q_{2}}}}F_{1}\left(q^{2}\right)
+\frac{p_{\alpha}p_{\nu}}{m_{\mathcal{B}_{Q_{1}Q_{2}}}^{2}}F_{2}\left(q^{2}\right)\right.\\ \notag
&&\left.+\frac{p_{\alpha}p^{\prime}_{\nu}}{m_{\mathcal{B}_{Q_{1}Q_{2}}}m_{\mathcal{B}_{Q_{1}}^{*}}}F_{3}\left(q^{2}\right)+g_{\alpha\nu}F_{4}\left(q^{2}\right)\right]\gamma_{5}u^{\mathcal{B}_{Q_{1}Q_{2}}}\left(p,s\right) \\ \notag
&&-\overline{u}_{\alpha}^{\mathcal{B}_{Q_{1}}^{*}}\left(p^{\prime},s^{\prime}\right)\left[\gamma_{\nu}\frac{p_{\alpha}}{m_{\mathcal{B}_{Q_{1}Q_{2}}}}G_{1}\left(q^{2}\right)
+\frac{p_{\alpha}p_{\nu}}{m_{\mathcal{B}_{Q_{1}Q_{2}}}^{2}}G_{2}\left(q^{2}\right)\right.\\
&&\left.+\frac{p_{\alpha}p^{\prime}_{\nu}}{m_{\mathcal{B}_{Q_{1}Q_{2}}}m_{\mathcal{B}_{Q_{1}}^{*}}}G_{3}\left(q^{2}\right)+g_{\alpha\nu}G_{4}\left(q^{2}\right)\right]u^{\mathcal{B}_{Q_{1}Q_{2}}}\left(p,s\right)
\end{eqnarray}
where $p$ and $p^{\prime}$ are the four-momentums of the mother and daughter particles, respectively. This process is realized by the transition $Q_{2}\rightarrow q^{\prime}l\overline{\nu}_{l}$ at quark level, where $q^{\prime}=u$ for $Q_{2}=b$, and $q^{\prime}=d/s$ for $Q_{2}=c$. $F_i/G_i$ $(i=1\sim4)$ are the vector and axial vector form factors.

Now, we introduce how these above form factors are obtained in the frame work of the QCD sum rules. We firstly write the three-point correlation function,
\begin{flalign}\label{eq:4}
\notag
&\Pi _{\mu\nu} (p,p')=& \\
& i^2\int d^4x d^4ye^{ip'x}e^{i(p-p')y} \left\langle 0 \right|T\{J_{\mu}^{\mathcal{B}_{Q_{1}}^{*}}(x)J^{V/A}_{\nu}(y)\bar J^{\mathcal{B}_{Q_{1}Q_{2}}}(0)\} \left| 0 \right\rangle &
\end{flalign}
where $T$ represents the time ordered product and $\bar{J}=J^\dagger\gamma_0$. $J^{\mathcal{B}_{Q_{1}Q_{2}}}$ and $J_{\mu}^{\mathcal{B}_{Q_{1}}^{*}}$ denote the interpolating currents of mother and daughter baryons, respectively. $J^{V/A}_\nu=\overline{q}^{\prime}\gamma_{\nu}(1-\gamma_{5})Q_{2}$ is the electroweak transition current. To extract the form factors in Eq. (\ref{eq:3}), this correlation function will be calculated at both hadron and quark levels, which are called the phenomenological side and the QCD side, respectively. Finally, we can extract the form factors by matching the calculation of these two sides.

At the hadronic level, we insert complete sets of hadronic states into the correlation function where all of the states that can couple to the interpolating currents are considered. After finishing the integration and using the double dispersion relation~\cite{Shifman:1978bx,Shifman:1978by}, the three-point correlation function is expressed as,
\begin{flalign}\label{eq:5}
\notag
&{\Pi _{\mu\nu}^{\mathrm{phy-V/A}} }(p,p')=& \\ \notag
&\frac{{\left\langle 0 \right|J_{\mu}^{\mathcal{B}_{Q_{1}}^{*}}\left| {\mathcal{B}^{*+}_{Q_{1}}} \right\rangle  \left\langle {\mathcal{B}^{*+}_{Q_{1}}} \right|J^{V/A}_{\nu}\left| {\mathcal{B}^{+}_{Q_{1}Q_{2}}} \right\rangle \left\langle {\mathcal{B}^{+}_{Q_{1}Q_{2}}} \right|{{\bar J}^{\mathcal{B}_{Q_{1}Q_{2}}}}\left| 0 \right\rangle }}{{(p{'^2} - m_{{\mathcal{B}^{*+}_{Q_{1}}}}^2)({p^2} - m_{\mathcal{B}_{Q_{1}Q_{2}}^{+}}^2)}}&
\\ \notag
&+\frac{{\left\langle 0 \right|J_{\mu}^{\mathcal{B}_{Q_{1}}^{*}}\left| {\mathcal{B}^{*-}_{Q_{1}}} \right\rangle  \left\langle {\mathcal{B}^{*-}_{Q_{1}}} \right|J^{V/A}_{\nu}\left| {\mathcal{B}^{+}_{Q_{1}Q_{2}}} \right\rangle \left\langle {\mathcal{B}^{+}_{Q_{1}Q_{2}}} \right|{{\bar J}^{\mathcal{B}_{Q_{1}Q_{2}}}}\left| 0 \right\rangle }}{{(p{'^2} - m_{{\mathcal{B}^{*-}_{Q_{1}}}}^2)({p^2} - m_{\mathcal{B}_{Q_{1}Q_{2}}^{+}}^2)}}&
\\ \notag
&+\frac{{\left\langle 0 \right|J_{\mu}^{\mathcal{B}_{Q_{1}}^{*}}\left| {\mathcal{B}^{*+}_{Q_{1}}} \right\rangle  \left\langle {\mathcal{B}^{*+}_{Q_{1}}} \right|J^{V/A}_{\nu}\left| {\mathcal{B}^{-}_{Q_{1}Q_{2}}} \right\rangle \left\langle {\mathcal{B}^{-}_{Q_{1}Q_{2}}} \right|{{\bar J}^{\mathcal{B}_{Q_{1}Q_{2}}}}\left| 0 \right\rangle }}{{(p{'^2} - m_{{\mathcal{B}^{*+}_{Q_{1}}}}^2)({p^2} - m_{\mathcal{B}_{Q_{1}Q_{2}}^{-}}^2)}}&
\\ \notag
&+\frac{{\left\langle 0 \right|J_{\mu}^{\mathcal{B}_{Q_{1}}^{*}}\left| {\mathcal{B}^{*-}_{Q_{1}}} \right\rangle  \left\langle {\mathcal{B}^{*-}_{Q_{1}}} \right|J^{V/A}_{\nu}\left| {\mathcal{B}^{-}_{Q_{1}Q_{2}}} \right\rangle \left\langle {\mathcal{B}^{-}_{Q_{1}Q_{2}}} \right|{{\bar J}^{\mathcal{B}_{Q_{1}Q_{2}}}}\left| 0 \right\rangle }}{{(p{'^2} - m_{{\mathcal{B}^{*-}_{Q_{1}}}}^2)({p^2} - m_{\mathcal{B}_{Q_{1}Q_{2}}^{-}}^2)}}&
\\ \notag
&+\frac{{\left\langle 0 \right|J_{\mu}^{\mathcal{B}_{Q_{1}}^{*}}\left| {\mathcal{B}^{+}_{Q_{1}}} \right\rangle  \left\langle {\mathcal{B}^{+}_{Q_{1}}} \right|J^{V/A}_{\nu}\left| {\mathcal{B}^{+}_{Q_{1}Q_{2}}} \right\rangle \left\langle {\mathcal{B}^{+}_{Q_{1}Q_{2}}} \right|{{\bar J}^{\mathcal{B}_{Q_{1}Q_{2}}}}\left| 0 \right\rangle }}{{(p{'^2} - m_{{\mathcal{B}^{+}_{Q_{1}}}}^2)({p^2} - m_{\mathcal{B}_{Q_{1}Q_{2}}^{+}}^2)}} & \\
\notag
&+\frac{{\left\langle 0 \right|J_{\mu}^{\mathcal{B}_{Q_{1}}^{*}}\left| {\mathcal{B}^{-}_{Q_{1}}} \right\rangle  \left\langle {\mathcal{B}^{-}_{Q_{1}}} \right|J^{V/A}_{\nu}\left| {\mathcal{B}^{+}_{Q_{1}Q_{2}}} \right\rangle \left\langle {\mathcal{B}^{+}_{Q_{1}Q_{2}}} \right|{{\bar J}^{\mathcal{B}_{Q_{1}Q_{2}}}}\left| 0 \right\rangle }}{{(p{'^2} - m_{{\mathcal{B}^{-}_{Q_{1}}}}^2)({p^2} - m_{\mathcal{B}_{Q_{1}Q_{2}}^{+}}^2)}} & \\
\notag
&+\frac{{\left\langle 0 \right|J_{\mu}^{\mathcal{B}_{Q_{1}}^{*}}\left| {\mathcal{B}^{+}_{Q_{1}}} \right\rangle  \left\langle {\mathcal{B}^{+}_{Q_{1}}} \right|J^{V/A}_{\nu}\left| {\mathcal{B}^{-}_{Q_{1}Q_{2}}} \right\rangle \left\langle {\mathcal{B}^{-}_{Q_{1}Q_{2}}} \right|{{\bar J}^{\mathcal{B}_{Q_{1}Q_{2}}}}\left| 0 \right\rangle }}{{(p{'^2} - m_{{\mathcal{B}^{+}_{Q_{1}}}}^2)({p^2} - m_{\mathcal{B}_{Q_{1}Q_{2}}^{-}}^2)}} & \\
\notag
&+\frac{{\left\langle 0 \right|J_{\mu}^{\mathcal{B}_{Q_{1}}^{*}}\left| {\mathcal{B}^{-}_{Q_{1}}} \right\rangle  \left\langle {\mathcal{B}^{-}_{Q_{1}}} \right|J^{V/A}_{\nu}\left| {\mathcal{B}^{-}_{Q_{1}Q_{2}}} \right\rangle \left\langle {\mathcal{B}^{-}_{Q_{1}Q_{2}}} \right|{{\bar J}^{\mathcal{B}_{Q_{1}Q_{2}}}}\left| 0 \right\rangle }}{{(p{'^2} - m_{{\mathcal{B}^{-}_{Q_{1}}}}^2)({p^2} - m_{\mathcal{B}_{Q_{1}Q_{2}}^{-}}^2)}} & \\
&+\cdots &
\end{flalign}
where the ellipsis denote the contributions of excited and continuum states. From this equation, we can see that the current $J^{\mathcal{B}_{Q_{1}Q_{2}}}$ can couple to the baryons $\mathcal{B}^{\pm}_{Q_{1}Q_{2}}$ with $J^{\texttt{P}}=\frac{1}{2}^{\pm}$, and the current $J^{\mathcal{B}_{Q_{1}}^{*}}$ can couple both to baryons $\mathcal{B}^{*\pm}_{Q_{1}}$ with $J^{\texttt{P}}=\frac{3}{2}^{\pm}$ and to $\mathcal{B}^{\pm}_{Q_{1}}$ with $J^{\texttt{P}}=\frac{1}{2}^{\pm}$. Each hadronic transition element $\left\langle {\mathcal{B}^{\pm}_{Q_{1}}/\mathcal{B}^{*\pm}_{Q_{1}}} \right|J^{V/A}_{\nu}\left| {\mathcal{B}^{\pm}_{Q_{1}Q_{2}}} \right\rangle$ in Eq. (\ref{eq:5}) can be parameterized as a series of vector/axialvector form factors similar as Eq. (\ref{eq:3}). To extract the form factors of the decay process $\frac{1}{2}^{+}\rightarrow\frac{3}{2}^{+}$, we must eliminate the interferences of the baryons with spin-parities $\frac{1}{2}^{\pm}$, $\frac{3}{2}^{-}$ in final states, and $\frac{1}{2}^{-}$ in initial states. As for the hadron vacuum matrix elements, they are defined as,
\begin{eqnarray}\label{eq:6}
\notag
&&\left\langle 0 \right|J_{\mu}^{\mathcal{B}_{Q_{1}}^{*}}(0)\left| \mathcal{B}_{Q_{1}}^{*+}(p^{\prime},s^{\prime}) \right\rangle  = \lambda_{\mathcal{B}_{Q_{1}}^{*+}}u_{\mu}(p^{\prime},s^{\prime})\\
\notag
&&\left\langle 0 \right|J_{\mu}^{\mathcal{B}_{Q_{1}}^{*}}(0)\left| \mathcal{B}_{Q_{1}}^{*-}(p^{\prime},s^{\prime}) \right\rangle  = \lambda_{\mathcal{B}_{Q_{1}}^{*-}}i\gamma_{5}u_{\mu}(p^{\prime},s^{\prime})\\
\notag
&&\left\langle 0 \right|J_{\mu}^{\mathcal{B}_{Q_{1}}^{*}}(0)\left| \mathcal{B}_{Q_{1}}^{+}(p^{\prime},s^{\prime}) \right\rangle  = \lambda_{\mathcal{B}_{Q_{1}}^{+}}i\gamma_{5}\Big(\alpha\gamma_{\mu}-\frac{4\alpha}{m}p^{\prime}_{\mu}\Big)u(p^{\prime},s^{\prime})\\
\notag
&&\left\langle 0 \right|J_{\mu}^{\mathcal{B}_{Q_{1}}^{*}}(0)\left| \mathcal{B}_{Q_{1}}^{-}(p^{\prime},s^{\prime}) \right\rangle  = \lambda_{\mathcal{B}_{Q_{1}}^{-}}\Big(\alpha\gamma_{\mu}-\frac{4\alpha}{m}p^{\prime}_{\mu}\Big)u(p^{\prime},s^{\prime})\\ \notag
&&\left\langle B_{Q_{1}Q_{2}}^{+}(p,s)\right|\overline{J}^{B_{Q_{1}Q_{2}}}\left|0\right\rangle=\lambda_{B_{Q_{1}Q_{2}}^{+}}\overline{u}(p,s)\\
&&\left\langle B_{Q_{1}Q_{2}}^{-}(p,s)\right|\overline{J}^{B_{Q_{1}Q_{2}}}\left|0\right\rangle=\lambda_{B_{Q_{1}Q_{2}}^{-}}\overline{u}(p,s)i\gamma_{5}
\end{eqnarray}
By substituting the matrix elements in Eq. (\ref{eq:5}) with these above relations, the correlation function on the phenomenological side can be decomposed into different dirac structures. It can be seen from Eq. (\ref{eq:6}) that the interferences of the current $J^{\mathcal{B}_{Q_{1}}^{*}}$ coupling to baryons $\mathcal{B}^{\pm}_{Q_{1}}$ with $J^{\texttt{P}}=\frac{1}{2}^{\pm}$ are related to the structures of $\gamma_{\mu}$ and $p_{\mu}^{\prime}$. Hence, the structures proportional to $\gamma_{\mu}$ and $p_{\mu}^{\prime}$ must be eliminated firstly. After these structures are eliminated, the correlation function can be decomposed into following different structures,
\begin{eqnarray}\label{eq:7}
\Pi_{\mu\nu}^{\mathrm{phy-V/A}}=\sum^{16}_{1}\Pi_{i}^{\mathrm{phy-V/A}}e_{i\mu\nu}^{V/A}
\end{eqnarray}
where $e_{i\mu\nu}^{V}$=$\gamma_{\nu}\gamma_{5}p_{\mu},\gamma_{\nu}\slashed p\gamma_{5}p_{\mu},\gamma_{\nu}\slashed p^{\prime}\gamma_{5}p_{\mu},\gamma_{\nu}\slashed p\slashed p^{\prime}\gamma_{5}p_{\mu},\gamma_{5}g_{\mu\nu},\\
\slashed p\gamma_{5}g_{\mu\nu},\slashed p^{\prime}\gamma_{5}g_{\mu\nu},\slashed p\slashed p^{\prime}\gamma_{5}g_{\mu\nu},\gamma_{5}p_{\mu}p_{\nu},\slashed p\gamma_{5}p_{\mu}p_{\nu},\slashed p^{\prime}\gamma_{5}p_{\mu}p_{\nu},\\
\slashed p\slashed p^{\prime}\gamma_{5}p_{\mu}p_{\nu},
\gamma_{5}p_{\mu}p^{\prime}_{\nu},\slashed p\gamma_{5}p_{\mu}p^{\prime}_{\nu},\slashed p^{\prime}\gamma_{5}p_{\mu}p^{\prime}_{\nu},
\slashed p\slashed p^{\prime}\gamma_{5}p_{\mu}p^{\prime}_{\nu}$
denote the structures about the vector part $\Pi_{i}^{\mathrm{phy-V}}$. The structures $e_{i\mu\nu}^{A}$ of axial vector part $\Pi_{i}^{\mathrm{phy-A}}$ can be obtained by removing $\gamma_{5}$ in these above structures. The form factors of the decaying process $\frac{1}{2}^{+}\rightarrow\frac{3}{2}^{+}$ are included in these expansion coefficients $\Pi^{\mathrm{phy-V/A}}_i$ which are commonly named as scalar invariant amplitudes.

In the following, let us concentrate on the QCD side where the interpolating currents in the correlation function in Eq. (\ref{eq:4}) are substituted with the following form,
\begin{eqnarray}\label{eq:10}
\notag
&&J^{\Xi^{\prime}_{Q_{1}Q_{2}}}= \frac{1}{\sqrt{2}}\varepsilon _{nml}\left({Q_{1}^{nT}\mathcal{C}\gamma _{\mu}Q_{2}^m}+{Q_{2}^{nT}\mathcal{C}\gamma _{\mu}Q_{1}^m}\right)\gamma _{\mu}\gamma _{5}q^l \\
\notag
&&J^{\Omega _{Q_{1}Q_{2}}}= \frac{1}{\sqrt{2}}\varepsilon _{nml}\left({Q_{1}^{nT}\mathcal{C}\gamma _{\mu}Q_{2}^m}+{Q_{2}^{nT}\mathcal{C}\gamma _{\mu}Q_{1}^m}\right)\gamma _{\mu}\gamma _{5}s^l \\
\notag
&&J^{\Sigma _{Q_{1}}^{*}}_{\mu}= \frac{1}{\sqrt{2}}\varepsilon _{nml}\left({u^{nT}\mathcal{C}\gamma _{\mu}d^m}+{d^{nT}\mathcal{C}\gamma _{\mu}u^m}\right)Q_{1}^l \\
\notag
&&J^{\Xi _{Q_{1}}^{\prime*}}_{\mu}= \frac{1}{\sqrt{2}}\varepsilon _{nml}\left({u^{nT}\mathcal{C}\gamma _{\mu}s^m}+{s^{nT}\mathcal{C}\gamma _{\mu}u^m}\right)Q_{1}^l \\
&&J^{\Omega _{Q_{1}}^{*}}_{\mu}= \varepsilon _{nml}\left({s^{nT}\mathcal{C}\gamma _{\mu}s^m}\right)Q_{1}^l
\end{eqnarray}
where $\varepsilon_{nml}$ is the 3 dimension Levi-Civita tensor, $n$, $m$ and $l$ represent the color indices, and $\mathcal{C}$ is the charge conjugation operator. Then, the correlation function is expressed as the following form after performing the Wick contraction,
\begin{flalign}\label{eq:11}
\notag
&\Pi_{\mu\nu} ^{\mathrm{QCD}}(p,p') =  2\sqrt{2}i^{2}\varepsilon _{nml}\varepsilon _{abc}\int d^4xd^4ye^{ip'x}e^{iqy}\Big\{ Q_{1}^{la}(x)\gamma _{\alpha} &\\
& \times \mathcal{C}Q_{2}^{ibT}(y)\mathcal{C}\gamma _\nu \mathcal{C}q_{3}^{miT}(x - y)\mathcal{C}\gamma _{\mu}q_{4}^{nc}(x)\gamma_{5}\gamma_{\alpha}\Big\} &
\end{flalign}
where $Q^{ij}(x)$ and $q^{ij}$ are the heavy and light quark full propagator. By doing the operator production expansion (OPE) and using the double dispersion relation, the correlation function on QCD side can be written as,
\begin{eqnarray}\label{eq:12}
\Pi _{\mu\nu}^{\mathrm{QCD}}(p,p') &&= \int\limits_{s_{\min}}^\infty  {ds} \int\limits_{u_{\min}}^\infty  {du} \frac{{\rho _{\mu\nu} ^{\mathrm{QCD}}(s,u,q^2)}}{{(s - p^2)(u - p'^2)}}
\end{eqnarray}
where $\rho^{\mathrm{QCD}}_{\mu\nu}(s,u,q^2)$ is QCD spectral density with $s=p^2$ and $u=p'^2$. $s_{\min}$ and $u_{\min}$ are create thresholds for initial and final baryons. The threshold is taken to be the square of the total masses of component quarks contained in the initial or final baryon.
Eliminating the structures proportional to $\gamma_{\mu}$ and $p_{\mu}^{\prime}$, the correlation function on QCD side can also be decomposed into the same structures as those on the phenomenological side,
\begin{eqnarray}\label{eq:13}
\Pi_{\mu\nu}^{\mathrm{QCD-V/A}}=\sum^{16}_{1}\Pi_{i}^{\mathrm{QCD-V/A}}e_{i\mu\nu}^{V/A}
\end{eqnarray}
The scalar invariant amplitudes on QCD side $\Pi^{\mathrm{QCD-V/A}}_i$ ($i=1,2\cdots16$) include perturbative part and different vacuum condensate terms $\langle{\bar qq}\rangle$, $\langle g_{s}^{2}GG\rangle$, $\langle \bar q g_{s}\sigma Gq\rangle$ and $g_{s}^{2}\langle{\bar qq}\rangle^{2}$.

\begin{figure}[htbp]
{\includegraphics[width=0.25\textwidth]{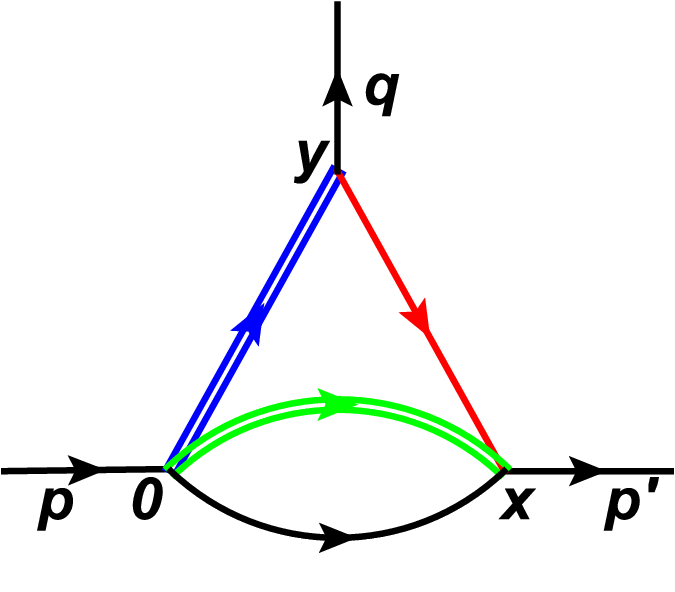}}
\caption{Feynman diagram of perturbative term at quark level with $q=p-p'$, where the double lines represent the heavy quark propagators and the single lines denote the propagators of u, d and s quarks.}
\label{pert}
\end{figure}

For the contribution of perturbative term as an example, we now introduce how the spectral density $\rho^{\mathrm{QCD-V}}_{\mu\nu,\mathrm{pert}}(s,u,q^2)$ is calculated. Firstly, the heavy and light quark propagators in Eq. (\ref{eq:11}) are substituted with the free propagators in momentum space. The corresponding Feynman diagram is shown as Fig. \ref{pert}. Its contribution can be written as,
\begin{flalign}\label{eq:14}
\notag
&\Pi _{\mu\nu,\mathrm{pert}}^{\mathrm{QCD-V}}(p,p') =- \frac{12\sqrt{2}}{{{{(2\pi )}^8}}}\int {{d^4}{k_1}} {d^4}{k_2}{d^4}{k_3}{d^4}{k_4}\int d^{4}q^{\prime}&\\ \notag
& \times\Big[\frac{{\delta ^4}(q^{\prime} - {k_1} - {k_4}) {\delta ^4}(p' - q^{\prime} - {k_3}){\delta ^4}(q - {k_2}+k_{3})}{{(k_{1}^{2}-m_{1}^{2})(k_{2}^{2}-m_{2}^{2})(k_{3}^{2}-m_{3}^{2})(k_{4}^{2} - m_{4}^{2})}} &\\
&\times({{\slashed k}_1}+m_{1})\gamma _{\alpha}({{\slashed k}_2}-m_{2}){\gamma _\nu }({{\slashed k}_3}-m_{3}){\gamma _\mu}({{\slashed k}_4} + {m_{4}}){\gamma _5}\gamma _{\alpha} \Big]&
\end{flalign}
By setting all quark lines on-shell with the Cutkosky's rule~\cite{Cutkosky}, its QCD spectral density function can be obtained as,
\begin{flalign}\label{eq:15}
\notag
&\rho _{\mu\nu,\mathrm{pert}} ^{\mathrm{QCD-V}}(s,u,{q^2})=-\frac{{12\sqrt{2}}}{{{{(2\pi )}^8}}}\frac{{{{( - 2\pi i)}^5}}}{{{{(2\pi i)}^3}}}&\\ \notag
& \times\int\limits_{(m_{1}+m_{4})^2}^{(\sqrt{u}-m_{3})^2} {dr^{\prime}} \int {{d^4}{k_1}}\delta (k_1^2-m_{1}^{2})\delta [{(q' - {k_1})^2}-m_{4}^{2}] & \\ \notag
&\times\int {{d^4}{k_3}} \delta [{(p' - {k_3})^2} - r^{\prime} ]\delta (k_3^2-m_{3}^{2})\delta [{({k_3} + p-p^{\prime})^2} - m_{2}^2]&\\
&\times({{\slashed k}_1}+m_{1})\gamma _{\alpha}({{\slashed k}_2}-m_{2}){\gamma _\nu }({{\slashed k}_3}-m_{3}){\gamma _\mu}({{\slashed k}_4} + {m_{4}}){\gamma _5}\gamma _{\alpha} \Big]&
\end{flalign}
where $q'=p'-k_{3}$, $k_{2}={k_3} + p-p^{\prime}$, $k_{4}=q'-k_{1}$ and $r'=q'^{2}$. These two phase-space-like integrals can be evaluated as,
\begin{flalign}\label{eq:16}
 \notag
&A(k_{1}^{2},m_1^2,m_4^2)=\int {{d^4}k_{1}\delta ({k_{1}^2} - m_1^2)\delta [{{(q' - k_{1})}^2} - m_4^2]}&\\
&= \frac{{\pi \sqrt {\lambda (r',m_1^2,m_4^2)} }}{{2r'}}&
\end{flalign}
\begin{flalign}\label{eq:17}
\notag
&B(k_{3}^{2},m_2^2,m_3^2)=&\\ \notag
&\int {{d^4}k_{3}} \delta ({k_{3}^2} - m_3^2)\delta [{(k_{3} + p-p')^2} - m_2^2]\delta [{(p' - k_{3})^2} - r']&\\
&= \frac{\pi }{{2\sqrt {\lambda (s,u,{q^2})} }}&
\end{flalign}
with the following constraint on the second phase space,
\begin{eqnarray}
\notag
\Big|\frac{{(u - {q^2} + m_2^2 - r')(s + u - {q^2}) + 2s(r'-u - m_3^2)}}{{\sqrt {{{(u - {q^2} + m_2^2 - r')}^2} - 4sm_3^2} \sqrt {\lambda (s,u,q^2)} }}\Big| \le 1
\end{eqnarray}
where $\lambda(a,b,c)=a^2+b^2+c^2-2(ab+ac+bc)$ is the triangle function. After performing the integrals of these above two phase space, we can obtain the QCD spectral density by setting $s=p^{2}$ and $u=p'^{2}$. As for the spectral densities of the vacuum condensate terms, they can also be calculated with the similar process as that of the perturbative term.

To improve the convergence in the quark-hadron duality and suppress the higher resonance and continuum state contributions, we usually perform double Borel transformations to both phenomenological and QCD sides. It is shown by Eqs. (\ref{eq:7}) and (\ref{eq:13}) that there are both 16 dirac structures at the phenomenological and QCD sides. Thus, we can establish 16 linear equations for the vector and axial vector form factors, respectively,
\begin{flalign}\label{eq:19}
 \notag
&\textbf{B}\textbf{B}_{i}\Big(F_{1}^{P_{1}P_{2}},F_{2}^{P_{1}P_{2}},F_{3}^{P_{1}P_{2}},F_{4}^{P_{1}P_{2}}\Big)
\times\texttt{exp}[-\frac{m_{\mathcal{B}_{Q_{1}Q_{2}}^{\texttt{P}_{1}}}^2}{M_{1}^{2}}-\frac{m_{\mathcal{B}_{Q_{1}}^{*\texttt{P}_{2}}}^2}{M_{2}^{2}}]e_{i\mu\nu}^{V}& \\ \notag
&=\int\limits_{s_{\min}}^{s_{0}}  {ds} \int\limits_{u_{\min}}^{u_{0}}  {du}\Big[\rho_{i} ^{\mathrm{QCD-V}}(s,u,Q^2)\times \texttt{exp}[-\frac{s}{M_{1}^{2}}-\frac{u}{M_{2}^{2}}]\Big]e_{i\mu\nu}^{V}& \\ \notag
 \notag
&\textbf{B}\textbf{B}_{i}\Big(G_{1}^{P_{1}P_{2}},G_{2}^{P_{1}P_{2}},G_{3}^{P_{1}P_{2}},G_{4}^{P_{1}P_{2}}\Big)
\times\texttt{exp}[-\frac{m_{\mathcal{B}_{Q_{1}Q_{2}}^{\texttt{P}_{1}}}^2}{M_{1}^{2}}-\frac{m_{\mathcal{B}_{Q_{1}}^{*\texttt{P}_{2}}}^2}{M_{2}^{2}}]e_{i\mu\nu}^{A}& \\
&=\int\limits_{s_{\min}}^{s_{0}}  {ds} \int\limits_{u_{\min}}^{u_{0}}  {du}\Big[\rho_{i} ^{\mathrm{QCD-A}}(s,u,Q^2)\times \texttt{exp}[-\frac{s}{M_{1}^{2}}-\frac{u}{M_{2}^{2}}]\Big]e_{i\mu\nu}^{A}&
\end{flalign}
with $i=1\sim16$ and $Q^2=-q^2$. Here, $M_{1}^{2}$ and $M_{2}^{2}$ are the Borel parameters and $\textbf{B}\textbf{B}_{i}\big(F_{1}^{P_{1}P_{2}}/G^{P_{1}P_{2}}_{1}\cdots F_{4}^{P_{1}P_{2}}/G^{P_{1}P_{2}}_{4}\big)$ denote the doubly Borel transformed coefficients on the phenomenological side. $F^{P_{1}P_{2}}_{j}/G^{P_{1}P_{2}}_{j}$ ($j=1\sim4$) are the form factors related to vector and axial vector transition current and the superscripts $P_{1}$ and $P_{2}$ denote the parities of the mother and daughter baryons. The spectral density functions $\rho_{i}^{\mathrm{QCD-V/A}}(s,u,Q^2)$ include contributions of perturbative part and different condensate terms. The full expressions of these functions are too complex to be shown here for simplicity. By solving these linear equations, we can extract the momentum dependent form factors $F_{i}(Q^2)$ and $G_{i}(Q^2)$. Here, $s_{0}$ and $u_{0}$ are threshold parameters which are used to eliminate the contributions of the excited and continuum states on the QCD side.
\section{Numerical results and Discussions}\label{sec3}
\begin{table*}[htbp]
\caption{1 Input parameters in this work. The masses of hadrons are in unit of GeV, and the values of pole residus of $\lambda$ are in units of GeV$^{3}$. The values of vacuum condensate are at the energy scale $\mu=1$ GeV.}
\begin{ruledtabular}
\renewcommand{\arraystretch}{1.3}
\label{PV}
\begin{tabular}{c c c c c c c c}
$m_{\Xi_{bc}(\frac{1}{2}^{+})}$&$m_{\Xi_{bc}(\frac{1}{2}^{-})}$&$m_{\Omega_{bc}(\frac{1}{2}^{+})}$&$m_{\Omega_{bc}(\frac{1}{2}^{-})}$ & $m_{\Omega_{bb}(\frac{1}{2}^{+})}$\\
$6.952$ \cite{Li:2022ywz}&$7.223$ \cite{Li:2022ywz}&$7.053$ \cite{Li:2022ywz}&$7.331$ \cite{Li:2022ywz}& $10.285$ \cite{Li:2022oth}&   &  \\ \hline
$m_{\Omega_{bb}(\frac{1}{2}^{-})}$&$m_{\Xi_{bb}(\frac{1}{2}^{+})}$&$m_{\Xi_{bb}(\frac{1}{2}^{-})}$&$m_{\Xi_{c}^{*\prime}(\frac{3}{2}^{+})}$&$m_{\Xi_{c}^{*\prime}(\frac{3}{2}^{-})}$ \\
 $10.528$ \cite{Li:2022oth}&$10.192$ \cite{Li:2022oth}& $10.428$ \cite{Li:2022oth}& $2.645$ \cite{Li:2022xtj}&$2.958$ \cite{Li:2022xtj}& &  \\ \hline
$m_{\Sigma_{c}^{*}(\frac{3}{2}^{+})}$&$m_{\Sigma_{c}^{*}(\frac{3}{2}^{-})}$&$m_{\Omega_{b}^{*}(\frac{3}{2}^{+})}$&$m_{\Omega_{b}^{*}(\frac{3}{2}^{-})}$&$m_{\Xi_{b}^{*\prime}(\frac{3}{2}^{+})}$ \\
$2.518$ \cite{Yu:2022ymb}&$2.829$ \cite{Yu:2022ymb}& $6.069$ \cite{Yu:2022ymb}& $6.336$ \cite{Yu:2022ymb}&$5.952$ \cite{Li:2022xtj}& &  \\ \hline
$m_{\Xi_{b}^{*\prime}(\frac{3}{2}^{-})}$&$m_{\Sigma_{b}^{*}(\frac{3}{2}^{+})}$&$m_{\Sigma_{b}^{*}(\frac{3}{2}^{-})}$&$\lambda_{\Xi_{bc}(\frac{1}{2}^{+})}$&$\lambda_{\Omega_{bc}(\frac{1}{2}^{+})}$ \\
$6.240$ \cite{Li:2022xtj}&$5.835$ \cite{Yu:2022ymb}& $6.116$ \cite{Yu:2022ymb}& $\sqrt{2}(0.125\pm0.016)$ \cite{Wang:2022ufh}&$\sqrt{2}(0.151\pm0.020)$ \cite{Wang:2022ufh}& &  \\ \hline
$\lambda_{\Xi_{bb}(\frac{1}{2}^{+})}$&$\lambda_{\Omega_{bb}(\frac{1}{2}^{+})}$ & $\lambda_{\Sigma_{c}^{*}(\frac{3}{2}^{+})}$& $\lambda_{\Xi_{c}^{*\prime}(\frac{3}{2}^{+})}$&	$\lambda_{\Omega_{b}^{*}(\frac{3}{2}^{+})}$	&    &     \\
$0.273$ \cite{Wang:2018lhz}&$0.327$ \cite{Wang:2018lhz}&$\sqrt{2}(0.027\pm0.008)$ \cite{Wang:2010vn}&$\sqrt{2}(0.033\pm0.008)$ \cite{Wang:2010vn}&$0.083\pm0.018$	\cite{Wang:2010vn}&   &      \\ \hline
$\lambda_{\Xi_{b}^{*\prime}(\frac{3}{2}^{+})}$  &  $\lambda_{\Sigma_{b}^{*}(\frac{3}{2}^{+})}$   & $\langle \overline{q}q\rangle$ & $\langle \overline{q}g_{s}\sigma Gq\rangle$ &$\langle \overline{s}s\rangle$         \\
$\sqrt{2}(0.049\pm0.012)$ \cite{Wang:2010vn}& $\sqrt{2}(0.038\pm0.011)$ \cite{Wang:2010vn} & $-(0.23\pm0.01)^{3}$ GeV$^{3}$ \cite{Shifman:1978by,Reinders}& $m_{0}^{2}\langle \overline{q}q\rangle$ \cite{ParticleDataGroup:2024cfk}&$(0.8\pm0.1)\langle \overline{q}q\rangle$ \cite{ParticleDataGroup:2024cfk}   \\ \hline
$\langle \overline{s}g_{s}\sigma Gs\rangle$& $\langle g_{s}^{2}GG\rangle$& $m_{0}^{2}$&  &     \\
$m_{0}^{2}\langle \overline{s}s\rangle$\cite{Shifman:1978by,Reinders} & $0.47\pm0.15$ GeV$^{4}$ \cite{Narison:2010cg,Narison:2011xe,Narison:2011rn}&$0.8\pm0.1$ GeV$^{2}$ \cite{Shifman:1978by,Reinders} &   & &   \\
\end{tabular}
\end{ruledtabular}
\end{table*}
The numerical results of QCDSR depend on some input parameters such as the masses of the heavy quarks and hadrons, and the values of vacuum condensates. The masses of heavy quarks and the values of vacuum condensates have dependence on energy scale. According to the renormlization group equation (RGE), this dependence can be expressed as,
\begin{flalign}\label{eq:20}
\notag
&m_{Q}(\mu )= m_{Q}(m_{Q})\left[\frac{\alpha _s(\mu )}{\alpha _s(m_{Q})}\right]^{\frac{12}{33 - 2n_f}}&\\
\notag
&m_{s}(\mu ) = m_{s}(2\mathrm{GeV})\left[\frac{\alpha _s(\mu )}{\alpha _s(2\mathrm{GeV})}\right]^{\frac{12}{33 - 2n_f}}&\\
\notag
&\left\langle \bar qq \right\rangle (\mu ) = \left\langle \bar qq \right\rangle (1\mathrm{GeV})\left[\frac{\alpha _s(1\mathrm{GeV})}{\alpha _s(\mu )}\right]^{\frac{12}{33 - 2n_f}}&\\
&\left\langle \bar qg_s\sigma Gq \right\rangle (\mu ) = \left\langle \bar qg_s\sigma Gq \right\rangle (1\mathrm{GeV})\left[\frac{\alpha _s(1\mathrm{GeV})}{\alpha _s(\mu )}\right]^{\frac{2}{33 - 2n_f}}&
\end{flalign}
\begin{flalign}\label{eq:21}
\notag
&\left\langle \bar sg_s\sigma Gs \right\rangle (\mu ) = \left\langle \bar sg_s\sigma Gs \right\rangle (1\mathrm{GeV})\left[\frac{\alpha _s(1\mathrm{GeV})}{\alpha _s(\mu )}\right]^{\frac{2}{33 - 2n_f}}&\\
&\alpha _s(\mu ) = \frac{1}{b_0t}\left[ 1 - \frac{b_1}{b_0^2}\frac{\log t}{t}  + \frac{b_1^2(\log ^2t - \log t - 1) + b_0b_2}{b_0^4t^2} \right]&
\end{flalign}
with $t=\log\frac{\mu^2}{\Lambda_{QCD}^2}$, $b_0=\frac{33-2n_f}{12\pi}$, $b_1=\frac{153-19n_f}{24\pi^2}$, $b_2=\frac{2857-\frac{5033}{9}n_f+\frac{325}{27}n_f^2}{128\pi^3}$, $\Lambda_{\mathrm{QCD}}=213$ MeV for the quark flavor $n_f=5$ in the present work~\cite{ParticleDataGroup:2024cfk}. The minimum subtraction masses of heavy quarks are adopted to be $m_c(m_c)=1.275\pm0.025$ GeV, $m_b(m_b)=4.18\pm0.03$ GeV and $m_s(2\mathrm{GeV})=0.095\pm0.005$ GeV \cite{ParticleDataGroup:2024cfk}. As for the energy scale, it is indicated that the value of $\mu=2$ GeV can work well in studying the properties of doubly heavy baryons containing bottom quarks~\cite{Wang:2022ufh}. The other parameters used in the present work are all listed in Tab. \ref{PV}.

\begin{figure}[htbp] \centering
\begin{subfigure}{\includegraphics[width=0.5\textwidth]{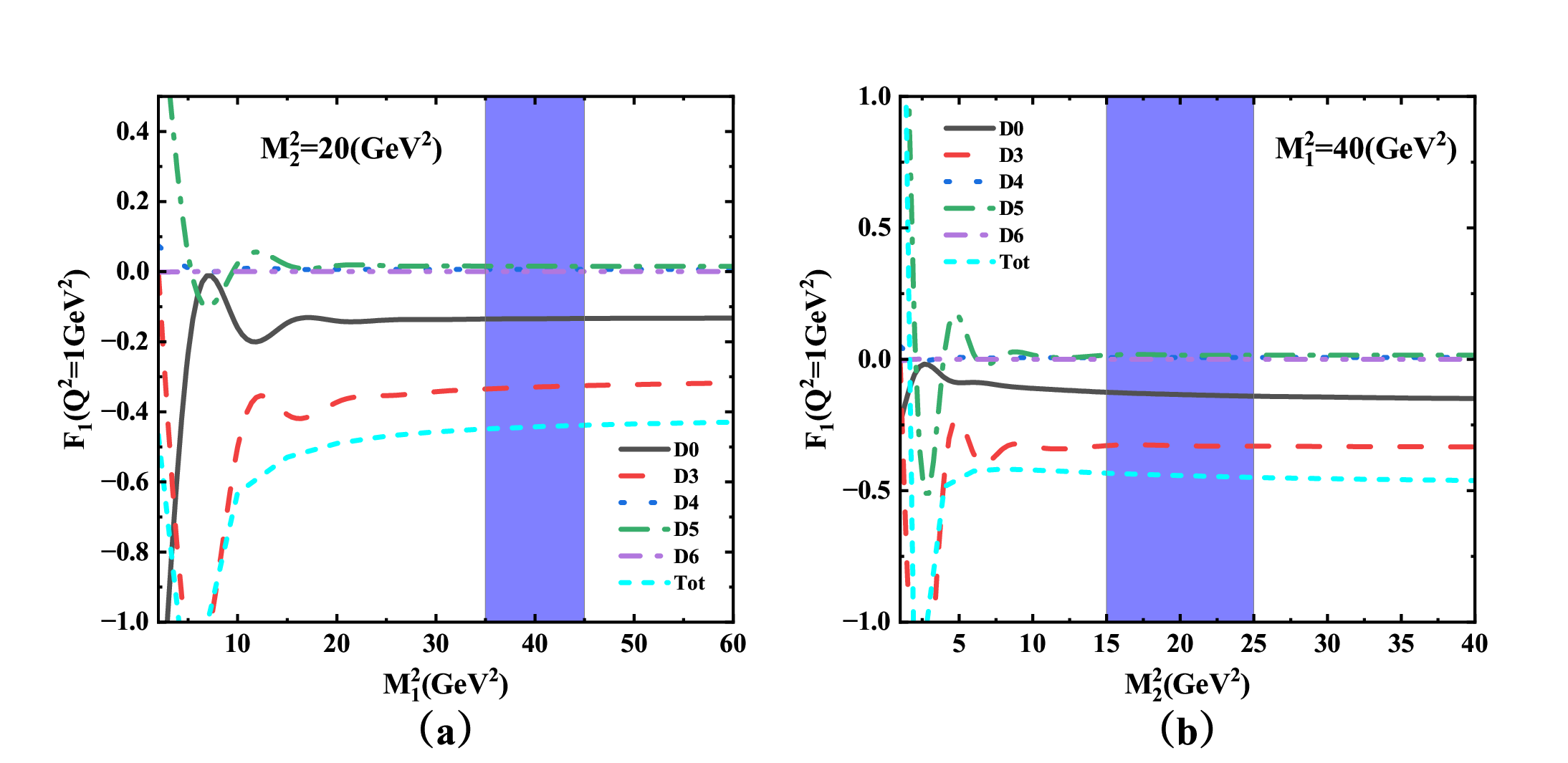}}
\caption{Contributions of perturbative term ($D_{0}$), different vacuum condensates ($D_{3}-D_{6}$) with respect to the
Borel parameter, where the grey area denotes the Borel platform. The results are for the $\Xi_{bc}^{+}\rightarrow\Sigma_{b}^{*0}$ transition process.}
\label{OPE}
\end{subfigure}
\begin{subfigure}{\includegraphics[width=0.5\textwidth]{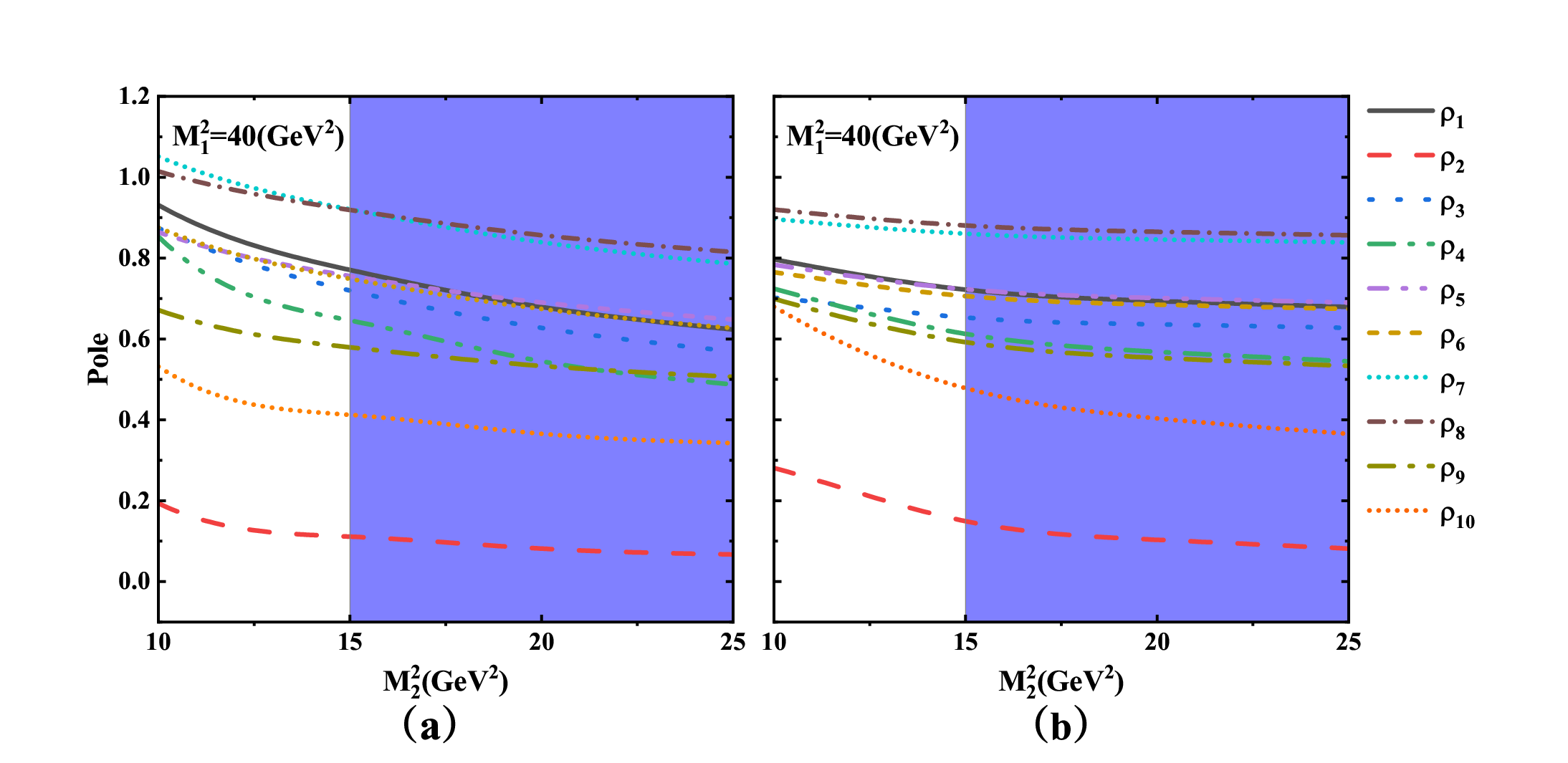}}
\caption{Pole contributions of \emph{s} and \emph{u} channel ((a) and (b)) for spectral densities $\rho_{1}-\rho_{10}$ with respect to the
Borel parameter $M_{2}^{2}$, where the grey area denotes the Borel platform. The results are for the $\Xi_{bc}^{+}\rightarrow\Sigma_{b}^{*0}$ transition process.}
\label{Pole}
\end{subfigure}
\end{figure}
From Eq. (\ref{eq:19}), we can see that the results of QCDSR also depend on the Borel parameters $M_{1}^{2}$, $M_{2}^{2}$ and the threshold parameters $s_0$, $u_0$. These parameters are determined according to searching for an appropriate working region named as 'Borel platform'. In this region, the values of form factors should have a weak dependence on the Borel parameters. At the same time, two conditions should be satisfied which are the pole dominance and the convergence of OPE. The pole contribution of $s$ and $u$ channels can be defined as \cite{Zhao:2020mod},
\begin{eqnarray}\label{eq:22}
\notag
\mathrm{Pole}_{s}=\frac{\int_{s_{min}}^{s_{0}} ds\int_{u_{min}}^{u_{0}} du\rho_{i}^{\mathrm{QCD}}\mathrm{exp}\left(-\frac{s}{M_{1}^{2}}-\frac{u}{M_{2}^{2}}\right)}{\int_{s_{min}}^{\infty} ds\int_{u_{min}}^{u_{0}} du\rho_{i}^{\mathrm{QCD}}\mathrm{exp}\left(-\frac{s}{M_{1}^{2}}-\frac{u}{M_{2}^{2}}\right)} \\
\mathrm{Pole}_{u}=\frac{\int_{s_{min}}^{s_{0}} ds\int_{u_{min}}^{u_{0}} du\rho_{i}^{\mathrm{QCD}}\mathrm{exp}\left(-\frac{s}{M_{1}^{2}}-\frac{u}{M_{2}^{2}}\right)}{\int_{s_{min}}^{s_{0}} ds\int_{u_{min}}^{\infty} du\rho_{i}^{\mathrm{QCD}}\mathrm{exp}\left(-\frac{s}{M_{1}^{2}}-\frac{u}{M_{2}^{2}}\right)}
\end{eqnarray}
\begin{figure}[htbp]
	\centering
	\includegraphics[width=8.5cm]{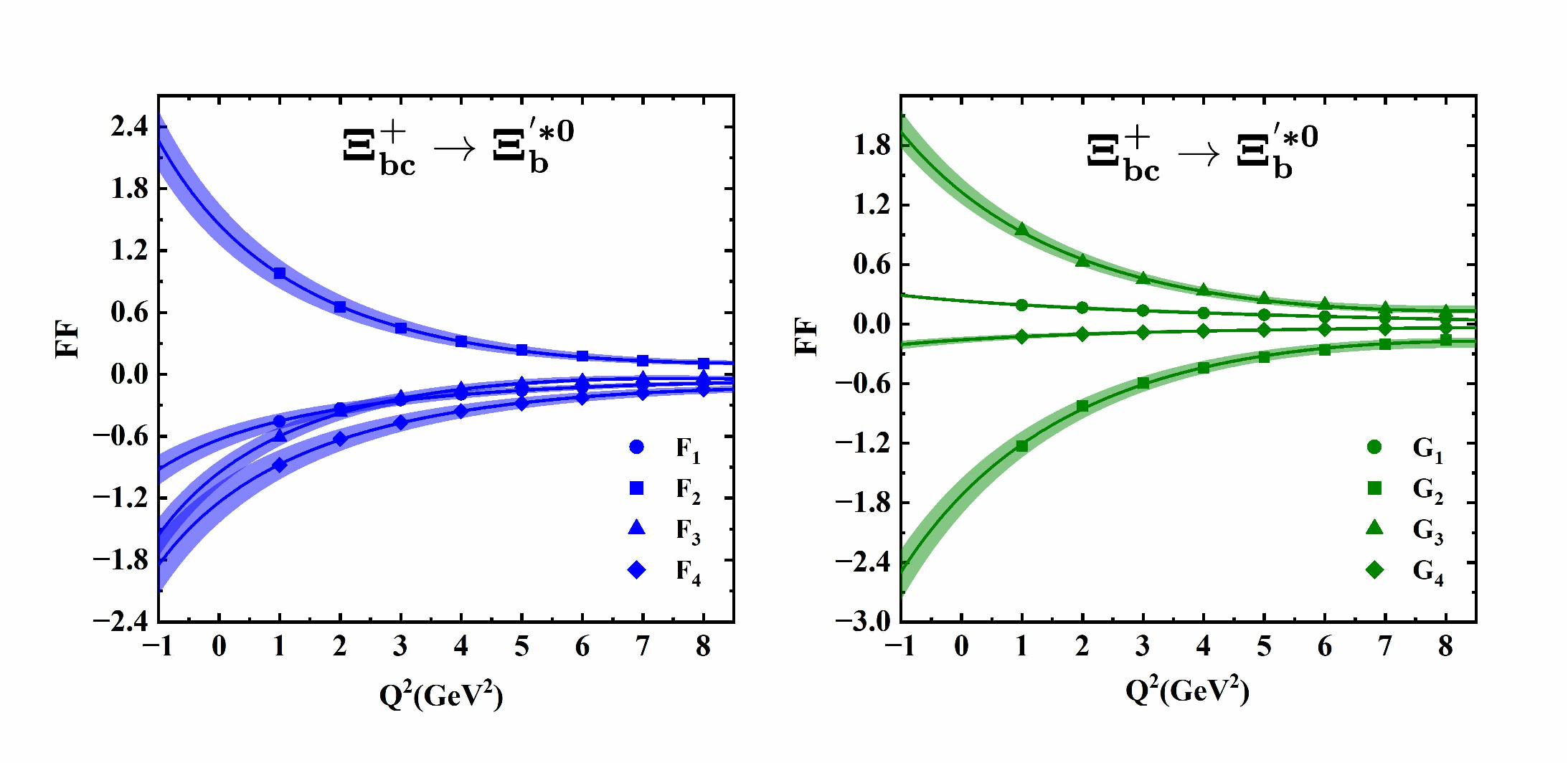}
	\centering
	\includegraphics[width=8.5cm]{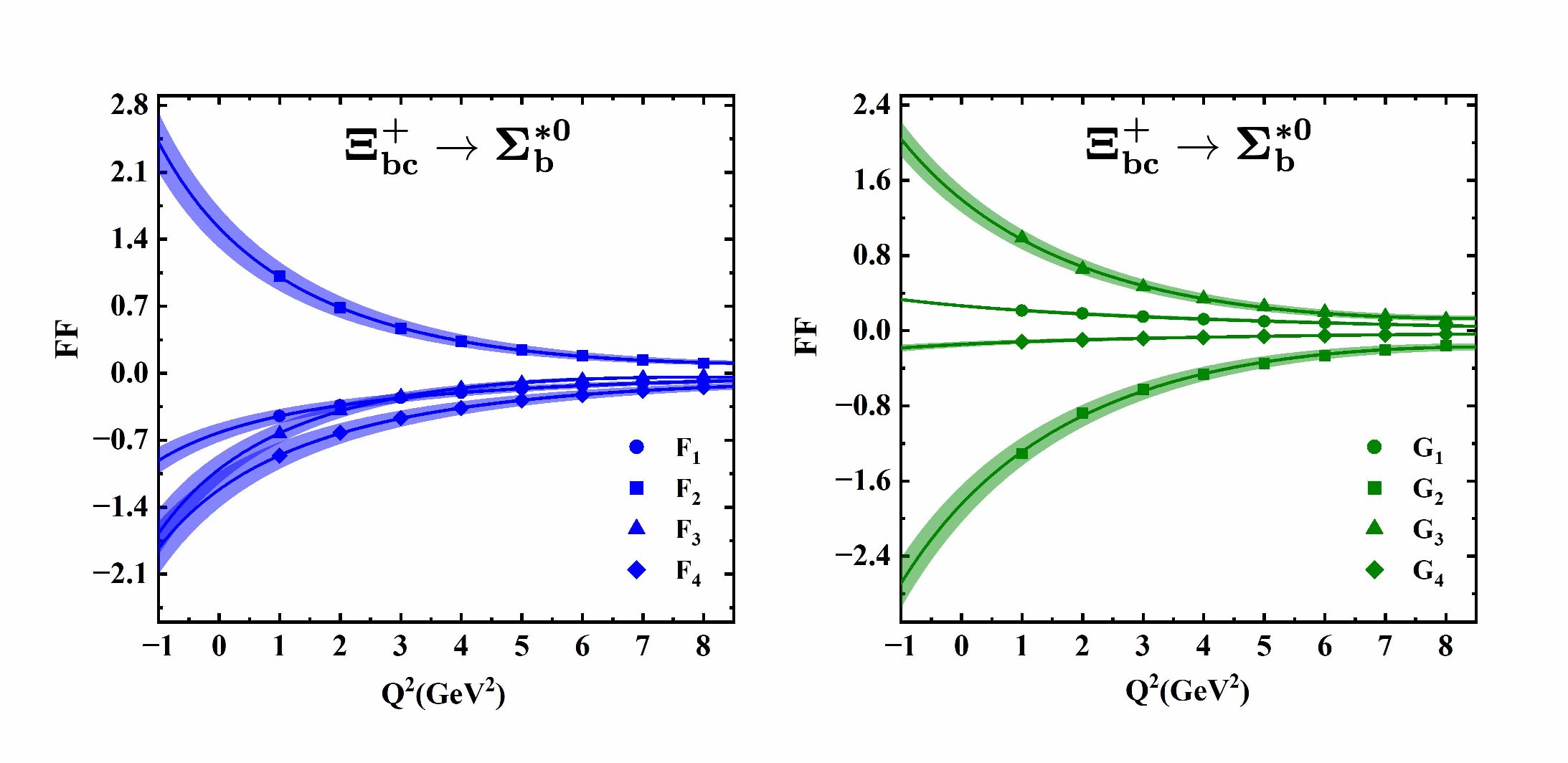}
	\caption{The fitting results of the form factors for transition processes $\Xi_{bc}^{+}\rightarrow\Xi_{b}^{\prime*0}$ and $\Xi_{bc}^{+}\rightarrow\Sigma_{b}^{*0}$.}
\label{fitting1}
\end{figure}
\begin{figure}[htbp]
	\centering
	\includegraphics[width=8.5cm]{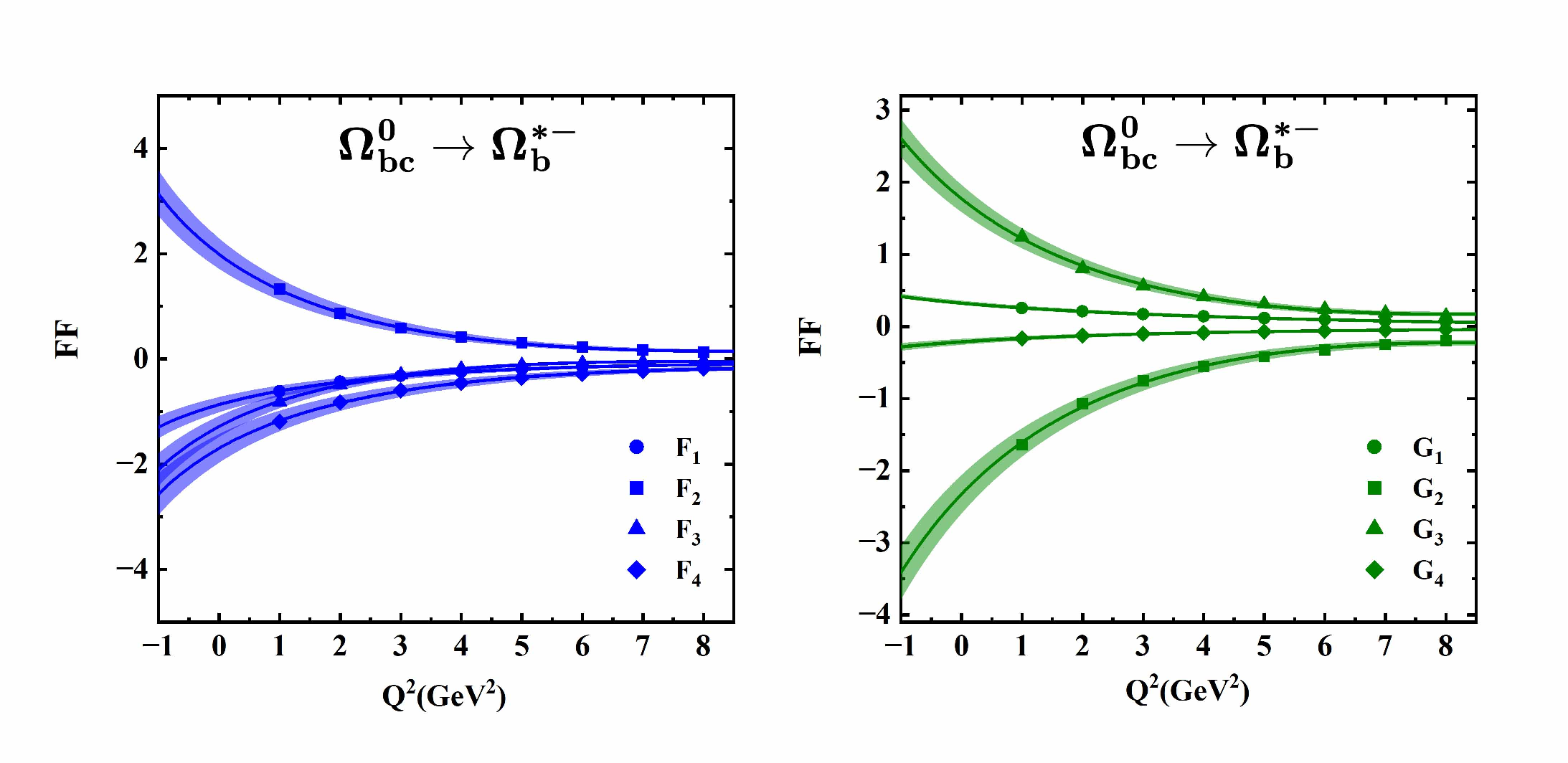}
	\centering
	\includegraphics[width=8.5cm]{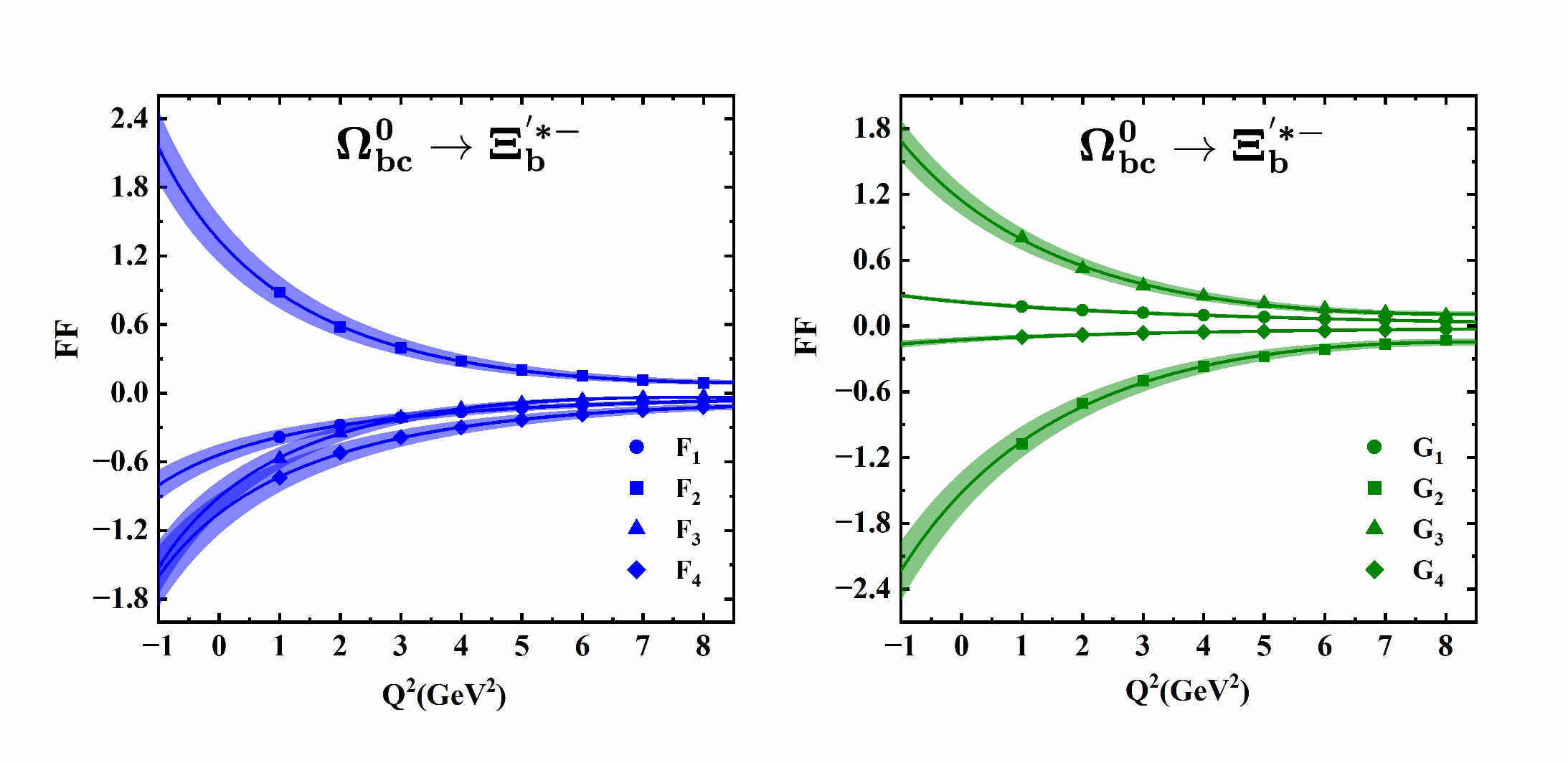}
	\caption{Same as Fig. 4 but for $\Omega_{bc}^{0}\rightarrow\Omega_{b}^{*-}$ and $\Omega_{bc}^{0}\rightarrow\Xi_{b}^{\prime*-}$.}
\label{fitting2}
\end{figure}
\begin{figure}[htbp]
	\centering
	\includegraphics[width=8.5cm]{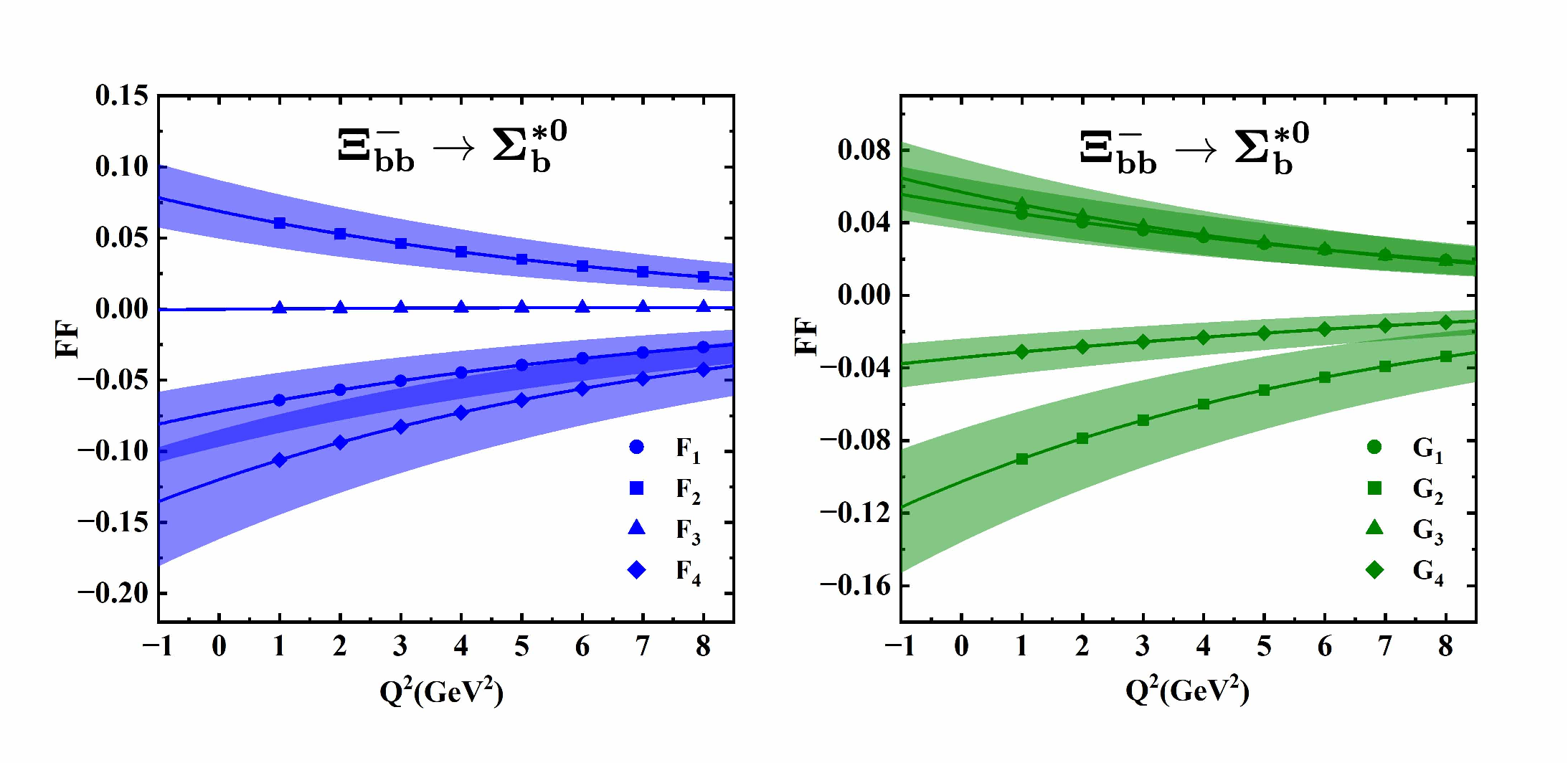}
	\centering
	\includegraphics[width=8.5cm]{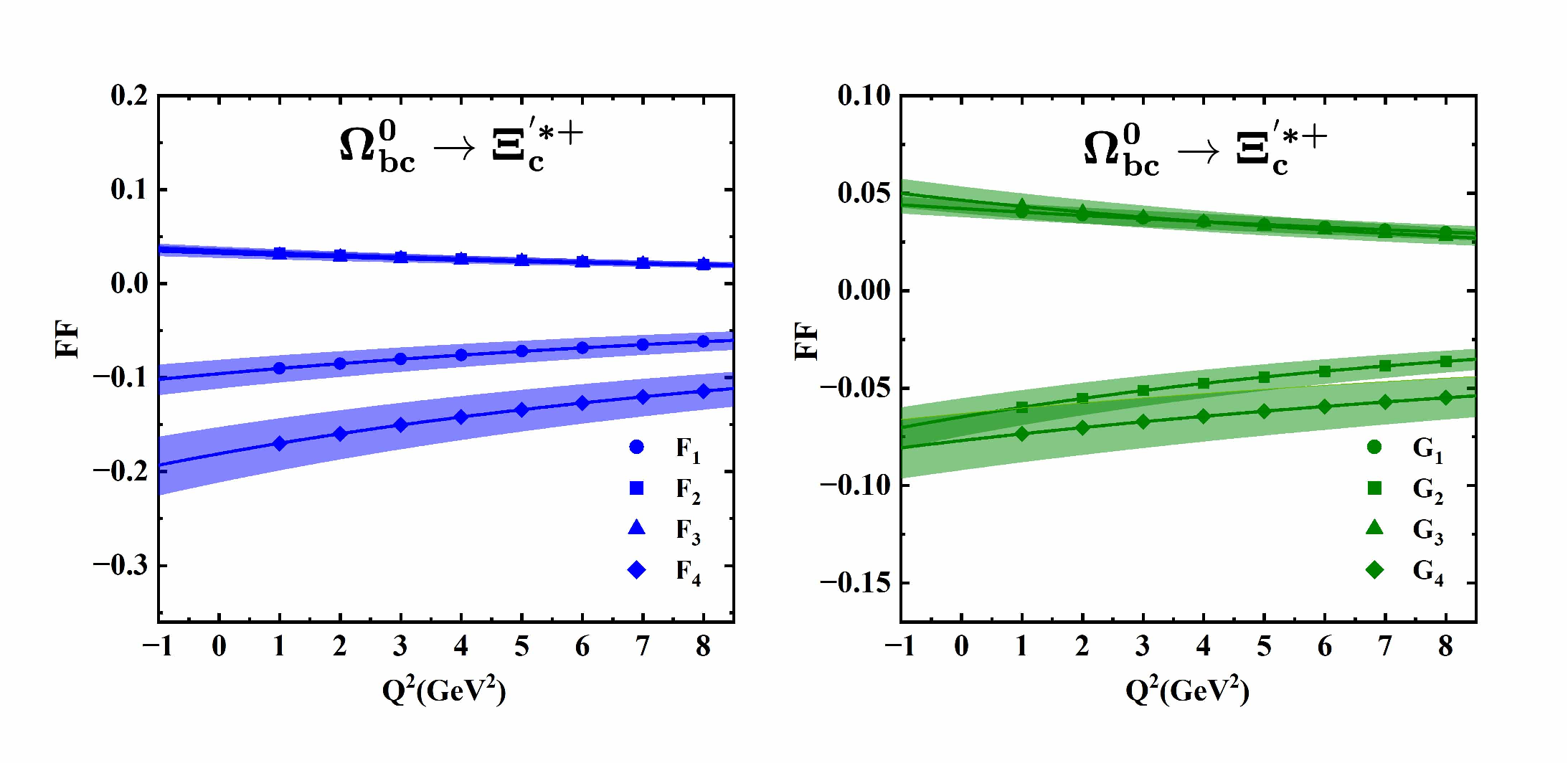}
	\caption{Same as Fig. 4 but for $\Xi_{bb}^{-}\rightarrow\Sigma_{b}^{*0}$ and $\Omega_{bc}^{0}\rightarrow\Xi_{c}^{\prime*+}$.}
\label{fitting3}
\end{figure}
\begin{figure}[htbp]
	\centering
	\includegraphics[width=8.5cm]{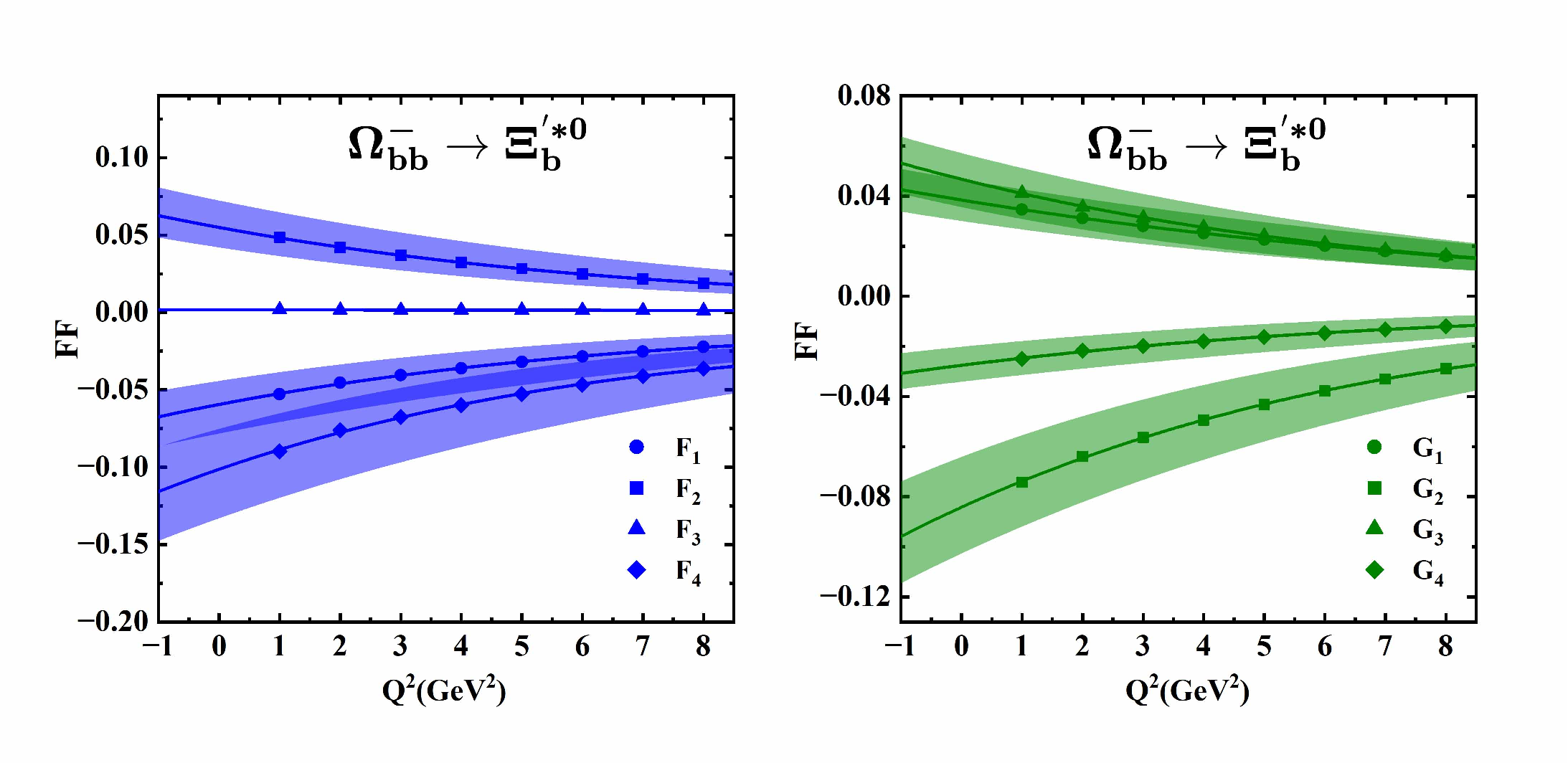}
	\centering
	\includegraphics[width=8.5cm]{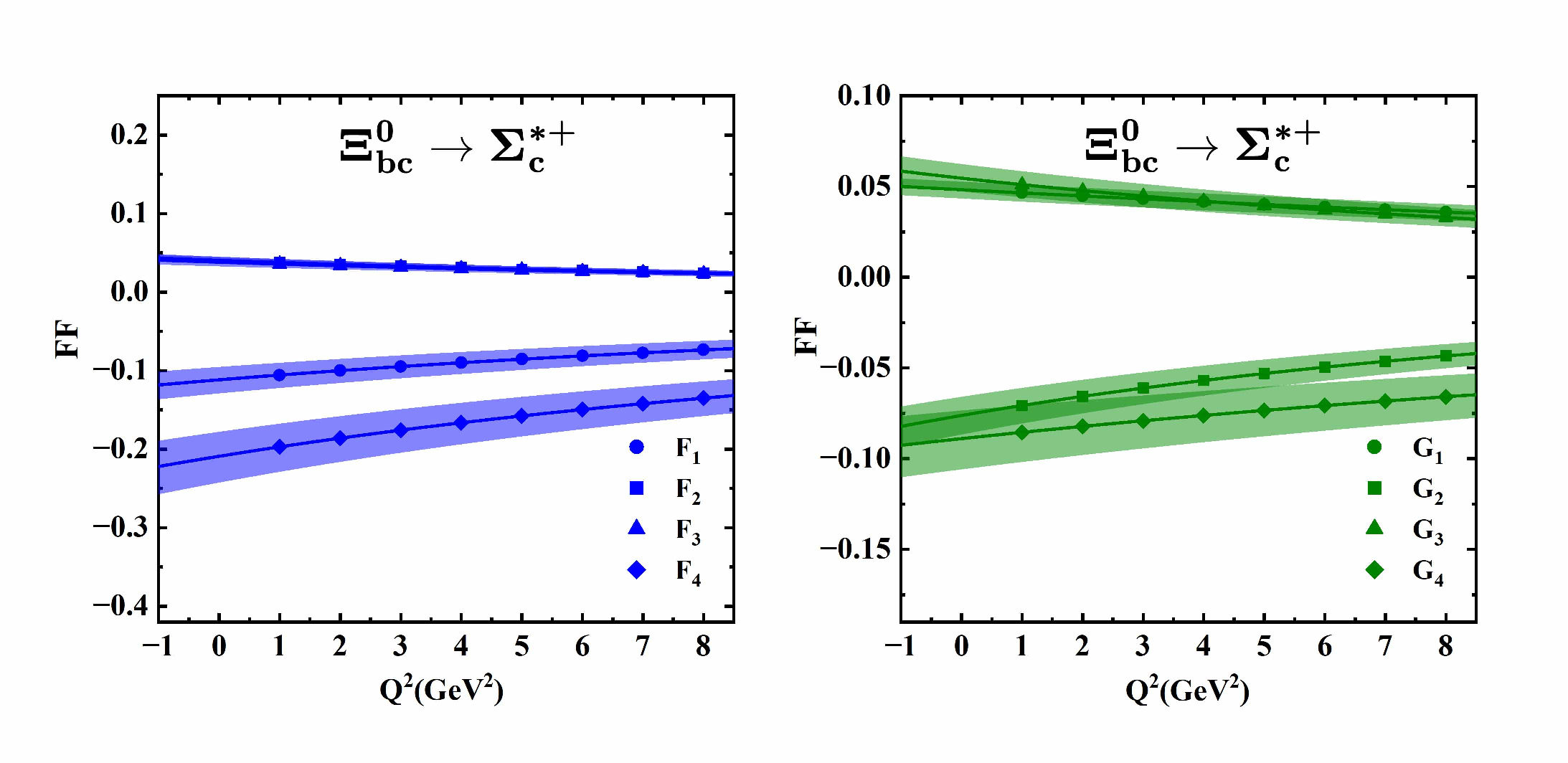}
	\caption{Same as Fig. 4 but for $\Omega_{bb}^{-}\rightarrow\Xi_{b}^{\prime*0}$ and $\Xi_{bc}^{0}\rightarrow\Sigma_{c}^{*+}$.}
\label{fitting4}
\end{figure}

The pole dominance requires that the pole contribution should be larger than $40\%$, and the convergence of OPE requires the contribution of high dimension condensate terms should be as small as possible. For the transition $\Xi_{bc}^{+}\rightarrow\Sigma_{b}^{*0}$ as an example, we plot the contributions of perturbative part and different vacuum condensate terms in Fig. \ref{OPE}. From this figure, we can see that main contributions come from quark condensate and purturbative part, and the larger the values of the Borel parameters, the smaller the contributions of vacuum condensate terms. The pole contributions of some spectral densities ($\rho_{1}\sim\rho_{10}$) with respect to $M_{2}^{2}$ are plotted in Fig. \ref{Pole}. After repeated trial and contrast, the Borel platform for $M_{2}^{2}$ is taken to be $15\sim25$ GeV$^2$ marked as blue area in Figs. \ref{OPE} (b) and \ref{Pole}. We can see that contributions of high dimension condensate terms ($D_{6}$) in the Borel platform are very small, which means the convergence of OPE is satisfied. Besides, Fig. \ref{Pole} shows that almost all of the pole contribution of $\rho_{1}\sim\rho_{10}$ are satisfied except for the spectral density of $\rho_{2}$ whose contribution is apparently lower than $40\%$. Actually, it is very difficult to guarantee all of dirac structures to satisfy the condition of pole dominance. As long as most of the structures satisfy this condition and an ideal Borel platform is found, reliable results can be extracted.
\begin{table*}[htbp]
\caption{The values of form factors (FF) at $Q^2=0$ for $c\rightarrow q$ transition process and fitting parameters by z series expand approach.}
\begin{ruledtabular}
\label{PR1}
\renewcommand{\arraystretch}{1.3}
\setlength{\tabcolsep}{1.5pt}
\footnotesize
\centering
\begin{tabular}{ccccccccccc}
Mode & FF &F(0) & $b_{0}$ & $b_{1}$ & $b_{2}$($\times10^{3})$  & FF &G(0) & $\tilde{b}_{0}$ & $\tilde{b}_{1}$ & $\tilde{b}_{2}$($\times10^{3})$   \\
\hline
\multirow{4}{*}{$\Omega_{bc}^{0}\rightarrow\Omega_{b}^{*-}$}
&$F_{1}$&$-0 .87_ {-0.15}^{+0.14}$&$-0.93_ {-0.15}^{+0.16}$ & $92 .80_ {-10.91}^{+12.90}$ & $-3.37_ {-0.35}^{+0.48}$
&$G_{1}$&$0 .32_ {-0.03}^{+0.03}$&$0 .33_ {-0.02}^{+0.03}$ & $-16.21_ {-2.04}^{+3.59}$ & $0 .03_ {-0.16}^{+0.26}$\\
&$F_{2}$&$1.99_ {-0.28}^{+0.30}$&$2 .19_ {-0.30}^{+0.32}$ & $-300.12_{-32.14}^{+32.98}$ & $12 .55_ {-1.14}^{+1.18}$
&$G_{2}$&$-2.33_ {-0.26}^{+0.27}$&$-2.57_ {-0.29}^{+0.29}$ & $357 .36_ {-29.19}^{+30.40}$ & $-15.47_ {-1.14}^{+1.22}$\\
&$F_{3}$&$-1.28_ {-0.20}^{+0.20}$&$-1.44_ {-0.22}^{+0.22}$ & $237 .38_ {-26.93}^{+29.75}$ & $-10.69_{-0.97}^{+1.18}$
&$G_{3}$&$1.77_ {-0.19}^{+0.19}$&$1 .96_ {-0.20}^{+0.21}$ & $-277.91_ {-22.00}^{+22.53}$ & $12 .19_ {-0.91}^{+0.95}$\\
&$F_{4}$&$-1.70_ {-0.27}^{+0.28}$&$-1.83_ {-0.29}^{+0.29}$ & $200 .11_ {-22.18}^{+27.59}$ & $-7.69_{-0.69}^{+1.09}$
&$G_{4}$&$-0.21_ {-0.04}^{+0.04}$&$-0.22_ {-0.04}^{+0.05}$ & $15 .95_ {-2.26}^{+2.35}$ & $-0.51_{-0.07}^{+0.08}$\\
\hline
\multirow{4}{*}{$\Omega_{bc}^{0}\rightarrow\Xi_{b}^{'*-}$}
&$F_{1}$&$-0.54_ {-0.10}^{+0.09}$&$-0.58_ {-0.10}^{+0.10}$ & $46 .31_ {-4.47}^{+6.30}$ & $-1.35_ {-0.08}^{+0.19}$
&$G_{1}$&$0.22_ {-0.02}^{+0.02}$&$0 .23_ {-0.02}^{+0.02}$ & $-8.09_ {-2.90}^{+3.24}$ & $-0.15_ {-0.18}^{+0.22}$\\
&$F_{2}$&$1.34_ {-0.20}^{+0.22}$&$1 .49_ {-0.21}^{+0.23}$ & $-189.94_ {-20.39}^{+20.42}$ & $7 .41_{-0.60}^{+0.66}$
&$G_{2}$&$-1.52_ {-0.19}^{+0.20}$&$-1.71_ {-0.21}^{+0.21}$ & $225 .26_ {-20.35}^{+21.16}$ & $-9.27_{-0.75}^{+0.81}$\\
&$F_{3}$&$-0.91_ {-0.14}^{+0.15}$&$-1.04_ {-0.16}^{+0.17}$ & $161 .52_ {-17.68}^{+19.97}$ & $-6.90_{-0.53}^{+0.72}$
&$G_{3}$&$1.15_ {-0.13}^{+0.14}$&$1 .29_ {-0.15}^{+0.15}$ & $-173.81_ {-14.94}^{+15.15}$ & $7 .27_{-0.59}^{+0.60}$\\
&$F_{4}$&$-1.05_ {-0.18}^{+0.18}$&$-1.14_ {-0.19}^{+0.19}$ & $105 .38_ {-12.66}^{+15.06}$ & $-3.49_{-0.37}^{+0.53}$
&$G_{4}$&$-0.13_ {-0.02}^{+0.03}$&$-0.13_ {-0.03}^{+0.03}$ & $6 .70_ {-1.06}^{+1.16}$ & $-0.14_{-0.03}^{+0.03}$\\
\hline
\multirow{4}{*}{$\Xi_{bc}^{+}\rightarrow\Sigma_{b}^{*0}$}
&$F_{1}$&$-0.62_ {-0.10}^{+0.10}$&$-0.66_ {-0.10}^{+0.10}$ & $43 .78_ {-2.12}^{+4.51}$ & $-0.94_ {-0.02}^{+0.11}$
&$G_{1}$&$0.27_ {-0.02}^{+0.01}$&$0 .27_ {-0.01}^{+0.01}$ & $-6.82_ {-5.35}^{+6.16}$ & $-0.36_ {-0.31}^{+0.32}$\\
&$F_{2}$&$1.52_ {-0.21}^{+0.21}$&$1 .71_ {-0.23}^{+0.23}$ & $-198.19_ {-13.63}^{+18.08}$ & $7 .05_ {-0.21}^{+0.50}$
&$G_{2}$&$-1.85_ {-0.20}^{+0.20}$&$-2.08_ {-0.21}^{+0.21}$ & $256 .89_ {-13.33}^{+15.72}$ & $-9.83_ {-0.37}^{+0.47}$\\
&$F_{3}$&$-1.00_ {-0.16}^{+0.15}$&$-1.16_ {-0.16}^{+0.18}$ & $170 .72_ {-14.44}^{+20.82}$ & $-6.93_ {-0.32}^{+0.72}$
&$G_{3}$&$1.40_ {-0.14}^{+0.14}$&$1 .58_ {-0.15}^{+0.15}$ & $-198.47_ {-10.36}^{+10.38}$ & $7 .71_ {-0.31}^{+0.34}$\\
&$F_{4}$&$-1.22_ {-0.19}^{+0.19}$&$-1.32_ {-0.20}^{+0.20}$ & $106 .30_ {-8.88}^{+12.68}$ & $-2.97_ {-0.20}^{+0.41}$
&$G_{4}$&$-1.15_ {-0.03}^{+0.03}$&$-0.15_ {-0.03}^{+0.03}$ & $4 .60_ {-0.62}^{+0.82}$ & $0 .01_ {-0.02}^{+0.03}$\\
\hline
\multirow{4}{*}{$\Xi_{bc}^{+}\rightarrow\Xi_{b}^{'*0}$}
&$F_{1}$&$-0.63_ {-0.11}^{+0.10}$&$-0.67_ {-0.11}^{+0.11}$ & $57 .02_ {-6.89}^{+7.08}$ & $-1.79_ {-0.23}^{+0.23}\times 10^{3}$
&$G_{1}$&$0.24_ {-0.01}^{+0.01}$&$0 .24_ {-0.01}^{+0.01}$ & $-5.50_ {-5.02}^{+6.17}$ & $-0.36_ {-0.31}^{+0.41}$\\
&$F_{2}$&$1.45_ {-0.20}^{+0.20}$&$1 .60_ {-0.21}^{+0.21}$ & $-202.32_ {-16.56}^{+18.79}$ & $7 .89_ {-0.43}^{+0.59}$
&$G_{2}$&$-1.72_ {-0.17}^{+0.19}$&$-1.90_ {-0.18}^{+0.21}$ & $241 .27_ {-13.22}^{+29.58}$ & $-9.67_ {-0.40}^{+1.99}$\\
&$F_{3}$&$-0.95_ {-0.12}^{+0.14}$&$-1.08_ {-0.12}^{+0.15}$ & $169 .74_ {-6.74}^{+14.78}$ & $-7.37_ {-0.24}^{+0.39}$
&$G_{3}$&$1.33_ {-0.12}^{+0.14}$&$1 .47_ {-0.13}^{+0.16}$ & $-191.04_ {-9.52}^{+24.90}$ & $7 .81_ {-0.32}^{+1.75}$\\
&$F_{4}$&$-1.24_ {-0.19}^{+0.20}$&$-1.33_ {-0.21}^{+0.21}$ & $128 .36_ {-15.97}^{+16.29}$ & $-4.43_ {-0.57}^{+0.58}$
&$G_{4}$&$-0.16_ {-0.03}^{+0.04}$&$-0.16_ {-0.03}^{+0.04}$ & $8 .98_ {-1.26}^{+2.84}$ & $-0.22_ {-0.04}^{+0.16}$
\end{tabular}
\end{ruledtabular}
\end{table*}
\begin{table*}[htbp]
\caption{The values of form factors (FF) at $Q^2=0$ for $b\rightarrow q$ transition process and fitting parameters by z series expand approach.}
\begin{ruledtabular}
\label{PR2}
\renewcommand{\arraystretch}{1.3}
\setlength{\tabcolsep}{1.5pt}
\footnotesize
\centering
\begin{tabular}{ccccccccccc}
Mode & FF &F(0) & $b_{0}$ & $b_{1}$ & $b_{2}$  & FF &G(0) & $\tilde{b}_{0}$ & $\tilde{b}_{1}$ & $\tilde{b}_{2}$   \\
\hline
\multirow{4}{*}{$\Omega_{bb}^{-}\rightarrow\Xi_{b}^{'*0}$}
&$F_{1}$&$-0.06_ {-0.02}^{+0.02}$&$-0.13_ {-0.004}^{+0.027}$ & $9 .65_ {-1.69}^{+2.74}$ & $-213.86_ {-36.61}^{+126.30}$
&$G_{1}$&$0.04_ {-0.01}^{+0.01}$&$0 .07_ {-0.004}^{+0.008}$ & $-4.03_ {-0.13}^{+0.84}$ & $59 .66_ {-15.59}^{+37.15}$\\
&$F_{2}$&$0.06_ {-0.02}^{+0.01}$&$0 .12_ {-0.014}^{+0.021}$ & $-9.03_ {-0.87}^{+1.07}$ & $190 .58_ {-15.40}^{+59.49}$
&$G_{2}$&$-0.08_ {-0.02}^{+0.02}$& $-0.19_ {-0.006}^{+0.033}$ & $14 .83_ {-1.69}^{+2.55}$ & $-322.85_ {-22.82}^{+122.48}$\\
&$F_{3}$&$0.002_ {-0.0004}^{+0.0002}$& $0 .001_ {-0.0001}^{+0.008}$ & $0 .11_ {-0.11}^{+1.30}$ & $-4.58_ {-4.02}^{+45.55}$
&$G_{3}$&$0.05_ {-0.01}^{+0.01}$& $0 .11_ {-0.004}^{+0.019}$ & $-8.18_ {-0.99}^{+1.32}$ & $178 .35_ {-14.77}^{+64.67}$\\
&$F_{4}$&$-0.10_ {-0.03}^{+0.03}$&$-0.23_ {-0.005}^{+0.051}$ & $18 .26_ {-3.42}^{+5.15}$ & $-423.45_ {-78.62}^{+237.26}$
&$G_{4}$&$-0.03_ {-0.007}^{+0.007}$&$-0.056_ {-0.004}^{+0.013}$ & $3 .87_ {-0.75}^{+1.76}$ & $-84.62_ {-17.23}^{+70.53}$\\
\hline
\multirow{4}{*}{$\Xi_{bb}^{-}\rightarrow\Sigma_{b}^{*0}$}
&$F_{1}$&$-0.07_ {-0.02}^{+0.02}$&$-0.14_ {-0.03}^{+0.04}$ & $8 .62_ {-1.06}^{+1.26}$ & $-136.94_ {-6.05}^{+11.65}$
&$G_{1}$&$0.05_ {-0.01}^{+0.01}$&$0 .09_ {-0.02}^{+0.02}$ & $-5.45_ {-0.07}^{+0.18}$ & $76 .97_ {-10.91}^{+15.83}$\\
&$F_{2}$&$0.07_ {-0.02}^{+0.02}$&$0 .15_ {-0.03}^{+0.03}$ & $-10.40_ {-1.17}^{+1.23}$ & $193 .24_ {-8.22}^{+9.68}$
&$G_{2}$&$-0.10_ {-0.03}^{+0.03}$&$-0.23_ {-0.05}^{+0.05}$ & $16 .75_ {-2.40}^{+2.45}$ & $-327.56_ {-31.44}^{+31.65}$\\
&$F_{3}$&$-0.0001_ {-0.0001}^{+0.0005}$&$-0.01_ {-0.00}^{+0.00}$ & $0 .83_ {-0.01}^{+0.06}$ & $-23.49_ {-0.41}^{+1.19}$
&$G_{3}$&$0.06_ {-0.02}^{+0.02}$&$0 .13_ {-0.03}^{+0.03}$ & $-9.20_ {-1.28}^{+1.29}$ & $179 .21_ {-15.16}^{+16.02}$\\
&$F_{4}$&$-0.12_ {-0.04}^{+0.04}$&$-0.25_ {-0.05}^{+0.06}$ & $15 .88_ {-2.31}^{+2.73}$ & $-273.77_ {-24.56}^{+34.51}$
&$G_{4}$&$-0.03_ {-0.01}^{+0.01}$&$-0.06_ {-0.01}^{+0.02}$ & $2 .83_ {-0.27}^{+0.33}$ & $-28.54_ {-0.68}^{+2.56}$\\
\hline
\multirow{4}{*}{$\Omega_{bc}^{0}\rightarrow\Xi_{c}^{'*+}$}
&$F_{1}$&$-0.10_ {-0.02}^{+0.01}$&$-0.13_ {-0.02}^{+0.02}$ & $1 .21_ {-0.10}^{+0.19}$ & $-4.80_ {-0.12}^{+1.00}$
&$G_{1}$&$0.04_ {-0.004}^{+0.004}$&$0 .05_ {-0.00}^{+0.00}$ & $-0.20_ {-0.03}^{+0.09}$ & $-0.49_ {-0.26}^{+0.83}$\\
&$F_{2}$&$0.04_ {-0.004}^{+0.004}$&$0 .05_ {-0.01}^{+0.01}$ & $-0.74_ {-0.04}^{+0.08}$ & $3 .72_ {-0.02}^{+0.44}$
&$G_{2}$&$-0.06_ {-0.01}^{+0.01}$&$-0.11_ {-0.01}^{+0.02}$ & $1 .92_ {-0.18}^{+0.33}$ & $-11.84_ {-0.61}^{+2.39}$\\
&$F_{3}$&$0.03_ {-0.01}^{+0.01}$&$0 .05_ {-0.01}^{+0.01}$ & $-0.59_ {-0.02}^{+0.14}$ & $3 .21_ {-0.35}^{+1.03}$
&$G_{3}$&$0.05_ {-0.01}^{+0.01}$&$0 .07_ {-0.01}^{+0.01}$ & $-1.09_ {-0.07}^{+0.19}$ & $6 .27_ {-0.05}^{+1.33}$\\
&$F_{4}$&$-0.18_ {-0.03}^{+0.03}$&$-0.25_ {-0.04}^{+0.04}$ & $2 .77_ {-0.24}^{+0.50}$ & $-13.43_ {-0.17}^{+2.99}$
&$G_{4}$&$-0.08_ {-0.01}^{+0.02}$&$-0.09_ {-0.02}^{+0.02}$ & $0 .58_ {-0.02}^{+0.10}$ & $-1.94_ {-0.38}^{+0.50}$\\
\hline
\multirow{4}{*}{$\Xi_{bc}^{0}\rightarrow\Sigma_{c}^{*+}$}
&$F_{1}$&$-0.11_ {-0.02}^{+0.02}$&$-0.14_ {-0.02}^{+0.02}$ & $0 .97_ {-0.10}^{+0.10}$ & $-1.84_ {-0.22}^{+0.26}$
&$G_{1}$&$0.05_ {-0.005}^{+0.004}$&$0 .05_ {-0.00}^{+0.00}$ & $0 .06_ {-0.10}^{+0.12}$ & $-2.66_ {-0.77}^{+0.82}$\\
&$F_{2}$&$0.04_ {-0.005}^{+0.005}$&$0 .06_ {-0.00}^{+0.01}$ & $-0.60_ {-0.01}^{+0.01}$ & $1 .85_ {-0.13}^{+0.14}$
&$G_{2}$&$-0.08_ {-0.01}^{+0.01}$&$-0.12_ {-0.01}^{+0.01}$ & $1 .77_ {-0.14}^{+0.14}$ & $-8.34_ {-0.45}^{+0.48}$\\
&$F_{3}$&$0.04_ {-0.01}^{+0.01}$&$0 .05_ {-0.01}^{+0.01}$ & $-0.52_ {-0.05}^{+0.06}$ & $1 .69_ {-0.12}^{+0.18}$
&$G_{3}$&$0.05_ {-0.01}^{+0.01}$&$0 .08_ {-0.01}^{+0.01}$ & $-1.02_ {-0.07}^{+0.08}$ & $4 .31_ {-0.13}^{+0.21}$\\
&$F_{4}$&$-0.21_ {-0.03}^{+0.03}$&$-0.27_ {-0.04}^{+0.04}$ & $2 .28_ {-0.26}^{+0.26}$ & $-6.83_ {-0.79}^{+0.90}$
&$G_{4}$&$-0.09_ {-0.02}^{+0.02}$&$-0.10_ {-0.02}^{+0.02}$ & $0 .27_ {-0.03}^{+0.03}$ & $0 .78_ {-0.07}^{+0.09}$\\
\end{tabular}
\end{ruledtabular}
\end{table*}

The threshold parameter $s_0$ and $u_0$ are used to eliminate the contributions of excited and continuum states, and their values are usually taken to be $s_0=(m_{\mathcal{B}_{Q_{1}Q_{2}}}+\Delta_{1})^{2}$ and $u_0=(m_{\mathcal{B}_{Q_{1}}^{*}}+\Delta_{2})^{2}$. It is shown by our previous work that the mass spectrum of the heavy baryons can be well determined by QCDSR with $\Delta_{1}=\Delta_{2}=0.6\sim0.7$ GeV \cite{Wang:2010vn}, thus the value of $\Delta_{1}=\Delta_{2}=0.6$ GeV are still adopted to analyze the form factors. After all of the parameters are determined, we plot variations of the form factors with respect to the Borel parameters $M_{1}^{2}$ and $M_{2}^{2}$ in Figs. \ref{Fig15}-\ref{Fig16} in Appendix \ref{Sec:AppA}. From these figures, we can see that the results show well stability and weak dependence on the Borel parameters in the Borel platform.

By taking different values of $Q^2$ in the range $1\sim8$ GeV$^2$, we can obtain the momentum dependent form factors $F_{i}(Q^2)$ and $G_{i}(Q^2)$ in space-like region ($Q^2>0$). The numerical results are explicitly shown in Figs. \ref{fitting1}-\ref{fitting4}. To obtain the values of the form factors in time-like region, these results will be fitted into appropriate analytical function and then extrapolated into time-like region ($Q^{2}<0$). A commonly used method to fit the form factors is the z series expand approach \cite{Boyd:1994tt} which can be expressed as,
\begin{eqnarray}\label{eq:23}
\notag
&&F_{i}(Q^2) = \\ \notag
&&\frac{1}{1+Q^2/m_{\mathrm{pole}}^{2}}\sum^{N-1}_{k=0}b_{k}\Big[z(Q^{2},t_{0})^{k}-(-1)^{k-N}\frac{k}{N}z(Q^{2},t_{0})^{N}\Big] \\
\notag
&&G_{i}(Q^2) = \\ \notag
&&\frac{1}{1+Q^2/m_{\mathrm{pole}}^{2}}\sum^{N-1}_{k=0}\widetilde{b}_{k}\Big[z(Q^{2},t_{0})^{k}-(-1)^{k-N}\frac{k}{N}z(Q^{2},t_{0})^{N}\Big] \\
\end{eqnarray}
with $z(Q^{2})=\frac{\sqrt{t_{+}+Q^{2}}-\sqrt{t_{+}-t_{0}}}{\sqrt{t_{+}+Q^{2}}+\sqrt{t_{+}-t_{0}}}$, $t_{\pm}=(m_{\mathcal{B}_{Q_{1}Q_{2}}}\pm m_{\mathcal{\mathcal{B}}_{2}^{*}})^{2}$ and $t_{0}=t_{+}-\sqrt{t_{+}(t_{+}-t_{-})}$.
In these equations, $b_{k}$ and $\widetilde{b}_{k}$ with $k=0\sim2$ are fitting parameters and their values are listed in Tabs. \ref{PR1} and \ref{PR2}. The fitting curves of the form factors are explicitly shown in Figs. \ref{fitting1}$\sim$\ref{fitting4}. From these figures, we can see that the form factors are fitted well by z series expand approach. Thus, it is reliable for us to obtain the values of form factors in time-like region which can be used to analyze the semileptonic decays of $\Xi_{bc}$, $\Omega_{bc}$, $\Xi_{bb}$ and $\Omega_{bb}$ baryons.
\begin{figure}[htbp]
	\centering
	\includegraphics[width=9cm]{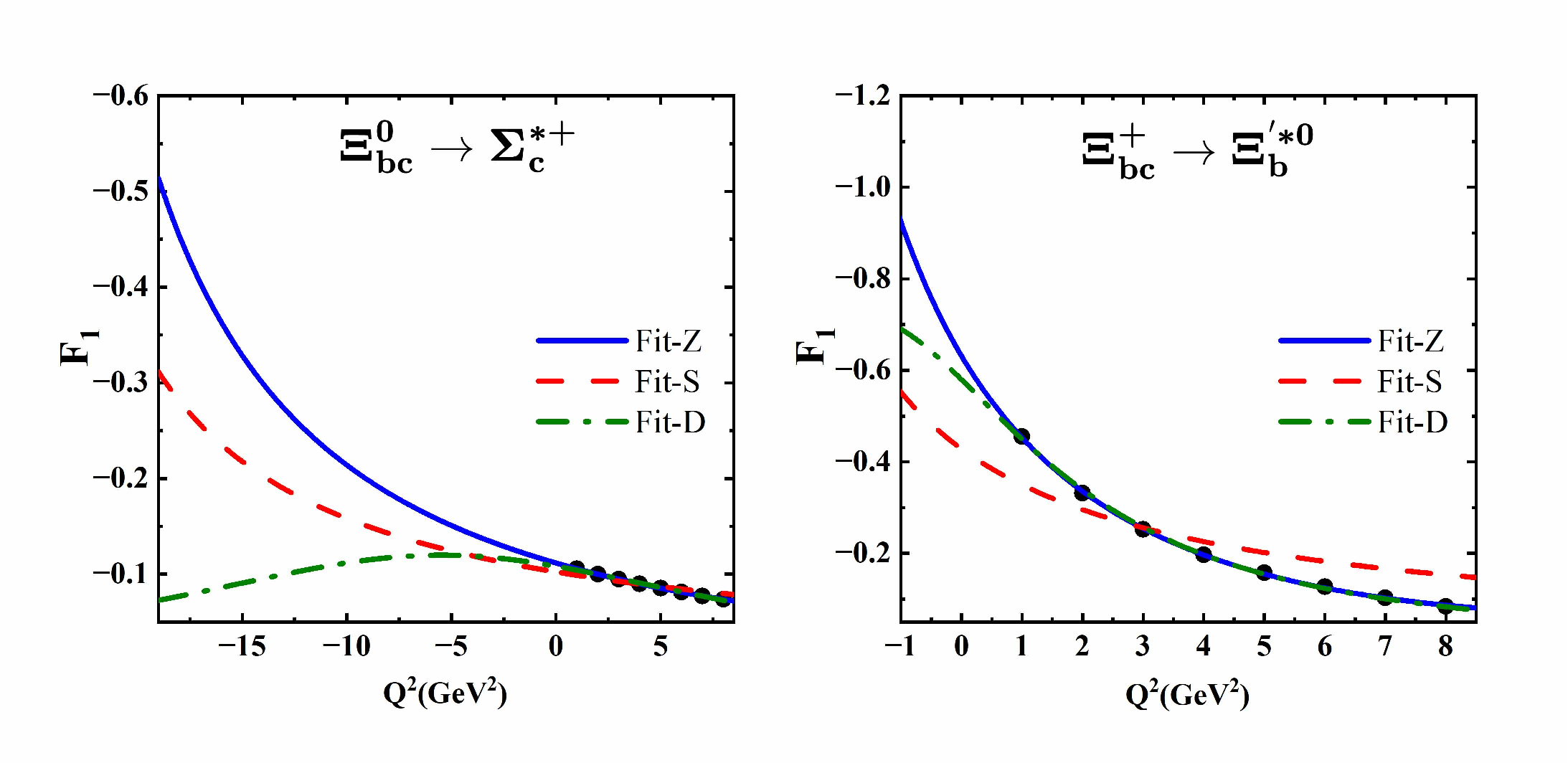}
		\caption{The fitting results of the form factor $F_{1}$ for $\Xi_{bc}^{0}\rightarrow\Sigma_{c}^{*+}$ and $\Xi_{bc}^{0}\rightarrow\Xi_{b}^{\prime*0}$ with different fitting methods, where Fit-Z, Fit-S and Fit-D denote the z series expand approach, single-pole structure and double-pole structure, respectively.}
\label{three fit}
\end{figure}

All of the values of the form factors $F_{i}(0)/G_{i}(0)$ are also listed in Tabs. \ref{PR1} and \ref{PR2}. It is noted that if the form factors of $F_{i}/G_{i}$ for transition processes $\Xi_{bb}^{-}\rightarrow\Sigma_{b}^{*0}$, $\Xi_{bc}^{0}\rightarrow\Sigma_{c}^{*+}$ and $\Xi_{bc}^{+}\rightarrow\Sigma_{b}^{*0}$ are multiplied a factor of $\sqrt{2}$, the values of form factors for $\Xi_{bb}^{0}\rightarrow\Sigma_{b}^{*+}$, $\Xi_{bc}^{+}\rightarrow\Sigma_{c}^{*++}$ and $\Xi_{bc}^{0}\rightarrow\Sigma_{b}^{*-}$ can also be obtained. Thus, the results of the latter are not listed in the table. In Refs. \cite{Zhao:2018mrg,Hu:2022xzu}, the similar research works were also carried out, where the light-front quark model and Light-Cone Sum Rules (LCSR) are employed, respectively. By comparison, we find that not all of the results are consistent well with each other. For the process $\Xi_{bc}^{0}\rightarrow \Sigma_{c}^{*+}$ as an example, our predicted central values for $F_{1}(0)\sim F_{4}(0)$ are ($-0.11$, $0.04$, $0.04$, $-0.21$) which is agreement well with the results ($-0.114$, $0.040$, $0.030$, $-0.239$) predicted by  quark model \cite{Zhao:2018mrg}, but have apparent discrepancy with the predictions ($0.04$, $-0.09$, $0.07$, $0.03$) by LCSR\cite{Hu:2022xzu}. Besides, the $F_{1}(0)\sim F_{4}(0)$ values of quark model for the transition $\Omega_{bc}^{0}\rightarrow\Omega_{b}^{*-}$ are predicted to be ($-2.465$, $14.850$, $-21.280$, $-4.494$)\cite{Zhao:2018mrg}, which are apparently higher than our results ($-0.87$, $1.99$, $-1.28$, $-1.70$).

Besides of employing different models, another difference between the present work and those of Refs. \cite{Zhao:2018mrg,Hu:2022xzu} is that they employ the single-pole and double-pole structures as their analytical functions to fit the numerical results of form factors. To illustrate the differences caused by the fitting functions, we plot the fitting results of $F_{1}$ for $\Xi_{bc}^{0}\rightarrow\Sigma_{c}^{*+}$ and $\Xi_{bc}^{0}\rightarrow\Xi_{b}^{\prime*0}$ processes with three fitting methods in Fig. \ref{three fit}. It is shown by this figure that the numerical results are more well fitted by the double-pole structure and the z series expand approach than the single-pole structure. Besides, the double-pole structure and the z series expand approach can lead to different dependence of the form factors on $Q^2$. As for the former, the values of form factors become lower after being extrapolated into the time-like region. However, the situation is exactly opposite for z series expand approach, where the values increase after being fitted into time-like region. Thus, we think the difference of the zero values $F_{i}(0)/G_{i}(0)$ between the present work and the others is related to both of the theoretical models and the fitting method.
\section{Semileptonic decays of $\mathcal{B}_{Q_{1}Q_{2}}(\frac{1}{2}^{+})\rightarrow\mathcal{B}_{Q_{1}}^{*}(\frac{3}{2}^{+})$}\label{sec4}
For the $\mathcal{B}_{Q_{1}Q_{2}}(\frac{1}{2}^{+})\rightarrow\mathcal{B}_{Q_{1}}^{*}(\frac{3}{2}^{+})$ weak transition process, its helicity amplitude can be expressed as,
\begin{eqnarray}\label{eq:24}
H_{\lambda_{2}\lambda_{W}}^{V,A}=\left\langle\mathcal{B}_{Q_{1}}^{*}\left(p^{\prime}\right)|\overline{q}^{\prime}\gamma_{\nu}(1-\gamma_{5})Q|\mathcal{B}_{Q_{1}Q_{2}}^{}\left(p\right) \right\rangle\epsilon^{*\nu}(\lambda_{W})
\end{eqnarray}
where $\lambda_{W}$ and $\lambda_{2}$ denote the helicity components of the $W$ boson and the daughter baryon $B_{Q_{1}}^{*}$ with their values being $\lambda_{W}$=$t$,$\pm1$,$0$ and $\lambda_{2}$=$\pm\frac{1}{2}$.
The relations of the helicity amplitudes to the form factors $F_{i}/G_{i}$ are explicitly shown in Ref. \cite{Faessler:2009xn}. The parametrization of the transition matrix element in the present work is different from Ref. \cite{Faessler:2009xn}, the relations between these form factors are as follows,
\begin{eqnarray}
\notag
&&F_{1}^{V}(q^{2})=F_{4}(q^{2}),\quad F_{1}^{A}(q^{2})=G_{4}(q^{2})\\ \notag
&&F_{2}^{V}(q^{2})=F_{1}(q^{2}),\quad F_{2}^{A}(q^{2})=G_{1}(q^{2}) \\ \notag
&&F^{V}_{3}(q^2)=\frac{m_{B_{1}}}{m_{B_{2}^{*}}}F_{3}(q^2)+F_{2}(q^{2}) \\ \notag
&&F^{A}_{3}(q^2)=\frac{m_{B_{1}}}{m_{B_{2}^{*}}}G_{3}(q^2)+G_{2}(q^{2})\\ \notag
&&F_{4}^{V}(q^{2})=F_{2}(q^{2}),\quad F_{4}^{A}(q^{2})=G_{2}(q^{2})
\end{eqnarray}

The differential decay of the semileptonic decay process for $\frac{1}{2}\rightarrow\frac{3}{2}$ can be expressed as,
\begin{eqnarray}\label{eq:25}
&&\frac{d\Gamma}{dq^{2}dcos\theta}=\frac{G_{F}^{2}|V_{CKM}|^{2}}{128\pi^{3}}\frac{(q^{2}-m_{l}^{2})}{m_{\mathcal{B}_{Q_{1}Q_{2}}}^{3}q^{2}}
\frac{\sqrt{Q_{+}Q_{-}}}{2}|M|^{2}
\end{eqnarray}
with $Q_{\pm}=(m_{\mathcal{B}_{Q_{1}Q_{2}}}\pm m_{\mathcal{B}_{Q_{1}}^{*}})^{2}-q^{2}$ and
\begin{widetext}
\begin{eqnarray}\label{eq:26}
\notag
|M|^{2}&&=L_{\mu\nu}H^{\mu\nu}\\ \notag
&&=\frac{2}{3}(q^2-m_l^2) \Bigg\{\frac{3}{8}(1\mp\cos\theta)^2\left(|H_{\frac{1}{2}1}|^2+|H_{\frac{3}{2}1}|^2\right)+
\frac{3}{8}(1\pm\cos\theta)^2\left(|H_{-\frac{1}{2}-1}|^2+|H_{-\frac{3}{2}-1}|^2\right)
+\frac{3}{4}\sin^2\theta\left(|H_{\frac{1}{2}0}|^2+|H_{-\frac{1}{2}0}|^2\right)\\ \notag
&&+\frac{m_l^2}{2q^2}\Bigg[\frac{3}{2}\left(|H_{\frac{1}{2}t}|^2+|H_{-\frac{1}{2}t}|^2\right)
+\frac{3}{4}\sin^2\theta\left(|H_{\frac{1}{2}1}|^2+|H_{-\frac{1}{2}-1}|^2+|H_{\frac{3}{2}1}|^2+|H_{-\frac{3}{2}-1}|^2\right)
+\frac{3}{2}\cos^2\theta\left(|H_{\frac{1}{2}0}|^2+|H_{-\frac{1}{2}0}|^2\right)\\
&&-3\cos\theta\left(H_{\frac{1}{2}t}H_{\frac{1}{2}0}+H_{-\frac{1}{2}t}H_{-\frac{1}{2}0}\right)\Bigg]\Bigg\}
\end{eqnarray}
\end{widetext}
where $L_{\mu\nu}$ and $H^{\mu\nu}$ are the leptonic and hadronic part. In the second line in Eq. (\ref{eq:26}), the upper and
lower signs refer to the configurations ($l^{-},\overline{\nu}_{l}$) and ($l^{+},\nu_{l}$), respectively and the same rule is also adopted in the following equations. $\theta$ is the polar angele of the lepton in the ($l$,$\nu_{l}$) c.m. system relative to the momentum direction of the $W$ boson. After finishing the integration to $\mathrm{cos}\theta$, the differential decay width can be decomposed into the following longitudinally and transversely polarized components,
\begin{eqnarray}\label{eq:27}
\notag
&&\frac{d\Gamma_{L}}{dq^{2}}=\frac{G_{F}^{2}|V_{CKM}|^{2}}{192\pi^{3}}\frac{(q^{2}-m_{l}^{2})^{2}}{m_{\mathcal{B}_{Q_{1}Q_{2}}}^{3}q^{2}}
\frac{\sqrt{Q_{+}Q_{-}}}{2}\Bigg[|H_{\frac{1}{2},0}|^{2}\\ \notag
&&+|H_{-\frac{1}{2},0}|^{2}+\frac{m_{l}^{2}}{2q^{2}}\Big(3|H_{\frac{1}{2},t}|^{2}+3|H_{-\frac{1}{2},t}|^{2}+|H_{\frac{1}{2},0}|^{2}+|H_{-\frac{1}{2},0}|^{2}\Big)\Bigg] \\
\end{eqnarray}
\begin{eqnarray}\label{eq:28}
\notag
&&\frac{d\Gamma_{T}}{dq^{2}}=\frac{G_{F}^{2}|V_{CKM}|^{2}}{192\pi^{3}}\frac{(q^{2}-m_{l}^{2})^{2}}{m_{\mathcal{B}_{Q_{1}Q_{2}}}^{3}q^{2}}
\frac{\sqrt{Q_{+}Q_{-}}}{2}\Bigg[|H_{\frac{1}{2},1}|^{2}\\ \notag
&&+|H_{-\frac{1}{2},-1}|^{2}+|H_{\frac{3}{2},1}|^{2}+|H_{-\frac{3}{2},-1}|^{2}+\frac{m_{l}^{2}}{2q^{2}}\Big(|H_{\frac{1}{2},1}|^{2}+|H_{-\frac{1}{2},-1}|^{2}\\
&&+|H_{\frac{3}{2},1}|^{2}+|H_{-\frac{3}{2},-1}|^{2}\Big)\Bigg]
\end{eqnarray}
The Fermi constant $G_{F}$, CKM matrix elements and masses of leptons are taken from the Particle Data Group \cite{ParticleDataGroup:2024cfk},
\begin{flalign}
\notag
&G_{F}=1.166\times10^{-5}\mathrm{GeV}^{-2}& \\ \notag
&|V_{cd}|=0.225,|V_{cs}|=0.974, |V_{ub}|=0.00357&\\ \notag
&m_{e}=0.511\times10^{-3}\mathrm{GeV},m_{\mu}=106\times10^{-3}\mathrm{GeV},m_{\tau}=1.78\mathrm{GeV}&
\end{flalign}
The dependence of differential decay widths on $q^{2}$ are explicitly shown in Figs. \ref{decay1} $\sim$ \ref{decay3} in Appendix \ref{Sec:AppB}.
By integrating out the square momentum $q^{2}$, the decay width can be written as,
\begin{eqnarray}\label{eq:29}
\Gamma=\int\limits_{m_{l}^{2}}^{(m_{\mathcal{B}_{Q_{1}Q_{2}}}-m_{\mathcal{B}_{Q_{1}}^{*}})^{2}}\Bigg\{\frac{d\Gamma_{L}}{dq^{2}}+\frac{d\Gamma_{T}}{dq^{2}}\Bigg\}dq^{2}
\end{eqnarray}
\begin{table*}[htbp]
\begin{ruledtabular}\caption{Results of semi-leptonic decay for $c\rightarrow q$ transition, where $\Gamma$ is in unit of $10^{-15}$ GeV and the lifetimes of $\Xi_{bc}^{+}$ and $\Omega_{bc}^{0}$ are taken to be 244 and 220 fs, respectively\cite{Karliner:2014gca,Kiselev:2001fw}.}
\label{result1}
\begin{tabular}{c| c c c c c c c c}
Modes& $\Gamma$&$\Gamma$\cite{Zhao:2018mrg}&$\Gamma$\cite{Hu:2022xzu}&$\mathcal{B}(10^{-3})$&$\Gamma_{L}/\Gamma_{T}$&${\langle A_{FB}^{l}\rangle}$&$\langle P_{z}^{F}\rangle$&$\langle P_{z}^{l}\rangle$\\ \hline
$\Xi_{bc}^{+}\rightarrow\Sigma_{b}^{*0}e^{+}\nu_{e}$&$0.32^{+0.16}_{-0.11}$&1.10 &  $0.19\pm0.04$  & $0.05^{+0.02}_{-0.02}$ & $0.87^{+0.06}_{-0.07}$&$0.31^{+0.02}_{-0.02}$&$0.43^{+0.03}_{-0.03}$&$1$\\
$\Xi_{bc}^{+}\rightarrow\Sigma_{b}^{*0}\mu^{+}\nu_{\mu}$&$0.31^{+0.15}_{-0.11}$&$-$ &  $0.19\pm0.04$ & $0.04^{+0.02}_{-0.02}$ & $0.87^{+0.06}_{-0.07}$&$0.29^{+0.02}_{-0.02}$&$0.43^{+0.03}_{-0.03}$&$0.94^{+0.003}_{-0.003}$\\
$\Omega_{bc}^{0}\rightarrow\Xi_{b}^{'*-}e^{+}\nu_{e}$&$0.19^{+0.10}_{-0.07}$& 0.70& $-$  & $0.06^{+0.03}_{-0.02}$ & $0.94^{+0.05}_{-0.06}$&$0.29^{+0.02}_{-0.01}$&$0.39_{-0.02}^{+0.02}$&$1$\\
$\Omega_{bc}^{0}\rightarrow\Xi_{b}^{'*-}\mu^{+}\nu_{\mu}$&$0.19^{+0.09}_{-0.07}$&$-$ & $-$  & $0.06^{+0.03}_{-0.02}$ & $0.93^{+0.05}_{-0.06}$&$0.26^{+0.02}_{-0.02}$&$0.39_{-0.02}^{+0.02}$&$0.93^{+0.002}_{-0.003}$\\
$\Xi_{bc}^{+}\rightarrow\Xi_{b}^{'*0}e^{+}\nu_{e}$&$3.86^{+2.02}_{-1.32}$&13.3 & $-$  &$0.55^{+0.29}_{-0.19}$ & $0.99^{+0.04}_{-0.05}$&$0.25^{+0.01}_{-0.02}$&$0.34^{+0.02}_{-0.02}$&$1$\\
$\Xi_{bc}^{+}\rightarrow\Xi_{b}^{'*0}\mu^{+}\nu_{\mu}$&$3.69^{+1.93}_{-1.26}$&$-$ & $-$  & $0.52^{+0.27}_{-0.18}$ & $0.98^{+0.04}_{-0.05}$&$0.22^{+0.01}_{-0.02}$&$0.34^{+0.02}_{-0.02}$&$0.92^{+0.002}_{-0.003}$\\
$\Omega_{bc}^{0}\rightarrow\Omega_{b}^{*-}e^{+}\nu_{e}$&$6.24^{+2.82}_{-2.12}$&14.7&  $-$   & $2.09^{+0.94}_{-0.71}$ & $1.02^{+0.03}_{-0.04}$&$0.24^{+0.01}_{-0.01}$&$0.33_{-0.015}^{+0.015}$&$1$\\
$\Omega_{bc}^{0}\rightarrow\Omega_{b}^{*-}\mu^{+}\nu_{\mu}$&$5.94^{+2.68}_{-2.02}$& $-$& $-$  & $1.99^{+0.90}_{-0.68}$ & $1.01^{+0.03}_{-0.04}$&$0.21^{+0.01}_{-0.01}$&$0.33_{-0.015}^{+0.015}$&$0.92^{+0.001}_{-0.002}$\\
\end{tabular}
\end{ruledtabular}
\end{table*}
\begin{table*}[htbp]
\begin{ruledtabular}\caption{Results of semi-leptonic decay for $b\rightarrow q$ transition, where $\Gamma$ is in unit of $10^{-17}$ GeV and the lifetimes of $\Omega_{bb}^{-}$, $\Xi_{bb}^{-}$ and $\Xi_{bc}^{0}$ are taken to be 800, 370 and 93 fs, respectively\cite{Karliner:2014gca,Kiselev:2001fw}.}
\label{result2}
\begin{tabular}{c| c c c c c c c c}
Modes& $\Gamma$& $\Gamma$\cite{Zhao:2018mrg}& $\Gamma$ \cite{Hu:2022xzu}&$\mathcal{B}(10^{-5})$&$\Gamma_{L}/\Gamma_{T}$&${\langle A_{FB}^{l}\rangle}$&$\langle P_{z}^{F}\rangle$&$\langle P_{z}^{l}\rangle$\\ \hline
$\Omega_{bb}^{-}\rightarrow\Xi_{b}^{'*0}e^{-}\overline{\nu}_{e}$&$1.09^{+0.34}_{-0.23}$& 1.90 & $-$ & $1.33^{+0.41}_{-0.28}$ & $0.38^{+0.03}_{-0.02}$&$-0.48^{+0.01}_{-0.03}$&$0.60^{+0.04}_{-0.01}$&$-1$\\
$\Omega_{bb}^{-}\rightarrow\Xi_{b}^{'*0}\mu^{-}\overline{\nu}_{\mu}$&$1.09^{+0.34}_{-0.23}$& $-$ & $-$&$1.32^{+0.41}_{-0.28}$ & $0.38^{+0.03}_{-0.02}$&$-0.48^{+0.01}_{-0.03}$&$0.60^{+0.04}_{-0.01}$&$-1$\\
$\Omega_{bb}^{-}\rightarrow\Xi_{b}^{'*0}\tau^{-}\overline{\nu}_{\tau}$&$0.66^{+0.25}_{-0.11}$& $-$ & $-$  &$0.81^{+0.31}_{-0.13}$ & $0.44^{+0.03}_{-0.02}$&$-0.35^{+0.02}_{-0.03}$&$0.56^{+0.04}_{-0.01}$&$-0.71^{+0.01}_{-0.008}$\\
$\Xi_{bb}^{-}\rightarrow\Sigma_{b}^{*0}e^{-}\overline{\nu}_{e}$&$1.45^{+0.83}_{-0.54}$& 1.94 & $7.72\pm1.80$  &$0.82^{+0.46}_{-0.30}$ & $0.33^{+0.02}_{-0.02}$&$-0.52^{+0.008}_{-0.004}$&$0.63_{-0.009}^{+0.003}$&$-1$\\
$\Xi_{bb}^{-}\rightarrow\Sigma_{b}^{*0}\mu^{-}\overline{\nu}_{\mu}$&$1.45^{+0.83}_{-0.54}$& $-$ & $7.72\pm1.80$  &$0.82^{+0.46}_{-0.30}$ & $0.33^{+0.02}_{-0.02}$&$-0.51^{+0.008}_{-0.004}$&$0.63_{-0.009}^{+0.003}$&$-1$\\
$\Xi_{bb}^{-}\rightarrow\Sigma_{b}^{*0}\tau^{-}\overline{\nu}_{\tau}$&$0.91^{+0.51}_{-0.33}$&  $-$ & $11.60\pm2.60$ &$0.51^{+0.28}_{-0.19}$ & $0.38^{+0.03}_{-0.02}$&$-0.39^{+0.012}_{-0.008}$&$0.60_{-0.008}^{+0.003}$&$-0.72^{+0.01}_{-0.01}$\\
$\Xi_{bc}^{0}\rightarrow\Sigma_{c}^{*+}e^{-}\overline{\nu}_{e}$&$2.48^{+1.00}_{-0.75}$&  1.66 & $1.47\pm0.32$  &$0.35^{+0.14}_{-0.11}$ & $0.29^{+0.05}_{-0.05}$&$-0.56^{+0.02}_{-0.02}$&$0.75^{+0.03}_{-0.03}$&$-1$\\
$\Xi_{bc}^{0}\rightarrow\Sigma_{c}^{*+}\mu^{-}\overline{\nu}_{\mu}$&$2.48^{+1.00}_{-0.75}$&  $-$ & $1.47\pm0.32$  &$0.35^{+0.14}_{-0.11}$ & $0.29^{+0.05}_{-0.05}$&$-0.56^{+0.02}_{-0.02}$&$0.75^{+0.03}_{-0.03}$&$-1$\\
$\Xi_{bc}^{0}\rightarrow\Sigma_{c}^{*+}\tau^{-}\overline{\nu}_{\tau}$&$1.45^{+0.58}_{-0.44}$& $-$ & $2.87\pm0.52$  &$0.20^{+0.08}_{-0.06}$ & $0.35^{+0.05}_{-0.05}$&$-0.42^{+0.03}_{-0.03}$&$0.72^{+0.02}_{-0.02}$&$-0.69^{+0.01}_{-0.01}$\\
$\Omega_{bc}^{0}\rightarrow\Xi_{c}^{'*+}e^{-}\overline{\nu}_{e}$&$2.09^{+0.80}_{-0.68}$&  1.26 & $-$  &$0.70^{+0.27}_{-0.23}$ & $0.34^{+0.04}_{-0.04}$&$-0.53^{+0.02}_{-0.02}$&$0.72_{-0.02}^{+0.02}$&$-1$\\
$\Omega_{bc}^{0}\rightarrow\Xi_{c}^{'*+}\mu^{-}\overline{\nu}_{\mu}$&$2.09^{+0.80}_{-0.68}$& $-$ & $-$  &$0.70^{+0.27}_{-0.23}$ & $0.34^{+0.04}_{-0.04}$&$-0.53^{+0.02}_{-0.02}$&$0.72_{-0.02}^{+0.02}$&$-1$\\
$\Omega_{bc}^{0}\rightarrow\Xi_{c}^{'*+}\tau^{-}\overline{\nu}_{\tau}$&$1.23^{+0.46}_{-0.40}$& $-$  & $-$ &$0.41^{+0.15}_{-0.13}$ & $0.39^{+0.04}_{-0.04}$&$-0.39^{+0.02}_{-0.02}$&$0.69_{-0.02}^{+0.02}$&$-0.69^{+0.01}_{-0.01}$																\end{tabular}
\end{ruledtabular}
\end{table*}

The integrated partial decay widths, ratios of $\Gamma_{L}/\Gamma_{T}$ and the corresponding branching fractions are calculated and the results are all listed in Tabs. \ref{result1} and \ref{result2}. For comparison, the results in Refs. \cite{Zhao:2018mrg,Hu:2022xzu} are also listed in the third and fourth columns in these tables. It is noted that the predicted form factors $F_{i}(0)/G_{i}(0)$ at present work are apparently lower than those of quark model. For $\Omega_{bc}^{0}\rightarrow\Omega_{b}^{*-}$ as an example, the $F_{i}(0)$ and $G_{i}(0)$ values are ($-2.465$, $14.850$, $-21.280$, $-4.494$) and ($-57.090$, $0.132$, $54.650$, $0.386$) in Ref \cite{Zhao:2018mrg}, which are apparently higher than our predictions ($-0.87$, $1.99$, $-1.28$, $-1.70$) and ($0.32$, $-2.33$, $1.77$, $-0.21$). However, the difference will become smaller after these numerical results are extrapolated into time-like region with two different fitting approaches. Thus, the decay widths for $\Omega_{bc}^{0}\rightarrow\Omega_{b}^{*-}e^{+}\nu_{e}$ predicted in present work are $6.24\times10^{-15}$ GeV which is comparable with the values of quark model $14.7\times10^{-15}$ GeV \cite{Zhao:2018mrg}.

Considering the flavor SU(3) symmetry, the studied semileptonic decay processes in the present work have the relations \cite{Wang:2017azm},
\begin{eqnarray}\label{eq:30}
\notag
&\Gamma\left(\Xi_{bc}^{+}\rightarrow\Sigma_{b}^{*0}l^{+}\nu\right)=\Gamma\left(\Omega_{bc}^{0}\rightarrow\Xi_{b}^{\prime*-}l^{+}\nu\right) \\ \notag
&\Gamma\left(\Xi_{bc}^{+}\rightarrow\Xi_{b}^{\prime*0}l^{+}\nu\right)=\frac{1}{2}\Gamma\left(\Omega_{bc}^{0}\rightarrow\Omega_{b}^{*-}l^{+}\nu\right)
\\
\notag
&\Gamma\left(\Xi_{bb}^{-}\rightarrow\Sigma_{b}^{*0}l^{-}\overline{\nu}\right)=\Gamma\left(\Omega_{bb}^{-}\rightarrow\Xi_{b}^{\prime*0}l^{-}\overline{\nu}\right) \\
&\Gamma\left(\Xi_{bc}^{0}\rightarrow\Xi_{c}^{\prime*+}l^{-}\overline{\nu}\right)=\Gamma\left(\Omega_{bc}^{0}\rightarrow\Xi_{c}^{\prime*+}l^{-}\overline{\nu}\right)
\end{eqnarray}
It is shown by the theoretical results in Tabs. \ref{result1} and \ref{result2} that these above relations are not strictly satisfied. For some of the decaying channels such as $c\rightarrow d/s$ transition processes listed in Tab. \ref{result1}, the SU(3) relation is significantly broken. This reflects the SU(3) symmetry breaking effecting which can be explained by the input parameters in the present work. Firstly, the masses of $u$ and $d$ quark is taken to be zero, and the mass of strange quark is not neglected (See Eq. (\ref{eq:20})). Secondly, the mass difference of $u/d$ and $s$ will will also lead to different masses of baryons such as $\Xi_{bc}^{+}$ and $\Omega_{bc}^{0}$.

With Eqs. (\ref{eq:25}) and (\ref{eq:26}), several physical observables in the semileptonic decay process can be evaluated such as the forward-backward asymmetry $A_{FB}^{l}$ of the lepton, the component $P_z^{F}$ of the polarization vector of final state baryons $\mathcal{B}_{Q_{1}}^{*}$ and the longitudinal polarization of the lepton $P_z^{l}$ for an unpolarized initial baryon. The forward-backward asymmetry of the lepton in the rest frame of $W$ boson is defined by \cite{Kadeer:2005aq},
\begin{eqnarray}\label{eq:31}
\notag
A_{FB}^{l}(q^2)&&=\frac{\frac{d\Gamma_{\text{forward}}}{dq^2}-\frac{d\Gamma_{\text{backward}}}{dq^2}}{\frac{d\Gamma}{dq^2}}\\ \notag
&&=\mp\frac{3}{4}
\Bigg\{\dfrac{\left(|H_{\frac{1}{2}1}|^2+|H_{\frac{3}{2}1}|^2\right)-\left(|H_{-\frac{1}{2}-1}|^2+|H_{-\frac{3}{2}-1}|^2\right)}{d\Gamma/dq^2}\\
&&-\dfrac{2m_l^2}{q^2}\dfrac{H_{\frac{1}{2}t}^*H_{\frac{1}{2}0}+H_{-\frac{1}{2}t}^*H_{-\frac{1}{2}0}}{d\Gamma/dq^2}\Bigg\}
\end{eqnarray}

The component $P_z^{F}(\theta)$ of the polarization vector is expressed as \cite{Kadeer:2005aq},
\begin{eqnarray}\label{eq:33}
P_z^{F}(\theta)=\frac{L_{\mu\nu}H_{++}^{\mu\nu}-L_{\mu\nu}H_{--}^{\mu\nu}}{L_{\mu\nu}H_{++}^{\mu\nu}+L_{\mu\nu}H_{--}^{\mu\nu}}
\end{eqnarray}
Here, $L_{\mu\nu}H_{--}^{\mu\nu}$ and $L_{\mu\nu}H_{++}^{\mu\nu}$ denote the contributions of positive and negative helicities of the daughter baryon and they can be written as,
\begin{eqnarray}\label{eq:34}
\notag
&&L_{\mu\nu}H_{--}^{\mu\nu}=\\ \notag
&&\frac{2}{3}(q^2-m_l^2)
\Bigg\{\frac{3}{8}(1\pm\cos\theta)^2\left(|H_{-\frac{1}{2}-1}|^2+|H_{-\frac{3}{2}-1}|^2\right)\\ \notag &&+\frac{3}{4}\sin^2\theta|H_{-\frac{1}{2}0}|^2+\frac{m_l^2}{2q^2}\Bigg[\frac{3}{2}|H_{-\frac{1}{2}t}|^2
+\frac{3}{4}\sin^2\theta\left(|H_{-\frac{1}{2}-1}|^2\right.\\
&&\left.+|H_{-\frac{3}{2}-1}|^2\right)+\frac{3}{2}\cos^2\theta|H_{-\frac{1}{2}0}|^2-3\cos\theta H_{-\frac{1}{2}t}H_{-\frac{1}{2}0}
\Bigg]\Bigg\}
\end{eqnarray}
\begin{eqnarray}\label{eq:35}
\notag
&&L_{\mu\nu}H_{++}^{\mu\nu}=\\ \notag
&&\frac{2}{3}(q^2-m_l^2)
\Bigg\{\frac{3}{8}(1\pm\cos\theta)^2\left(|H_{\frac{1}{2}1}|^2+|H_{\frac{3}{2}1}|^2\right)\\ \notag &&+\frac{3}{4}\sin^2\theta|H_{\frac{1}{2}0}|^2+\frac{m_l^2}{2q^2}\Bigg[\frac{3}{2}|H_{\frac{1}{2}t}|^2
+\frac{3}{4}\sin^2\theta\left(|H_{\frac{1}{2}1}|^2\right.\\
&&\left.+|H_{\frac{3}{2}1}|^2\right)+\frac{3}{2}\cos^2\theta|H_{\frac{1}{2}0}|^2-3\cos\theta H_{\frac{1}{2}t}H_{\frac{1}{2}0}
\Bigg]\Bigg\}
\end{eqnarray}
The longitudinal polarization $P_z^{l}$ of the lepton for the decay of an unpolarized initial baryon is \cite{Kadeer:2005aq},
\begin{eqnarray}\label{eq:36}
P_z^{l}(\theta)=\pm\frac{L_{\mu\nu}H^{\mu\nu}(flip)-L_{\mu\nu}H^{\mu\nu}(nonflip)}{L_{\mu\nu}H^{\mu\nu}(flip)+L_{\mu\nu}H^{\mu\nu}(nonflip)}
\end{eqnarray}
Here, the flip transition part is related to the factor of $\frac{m_{l}^{2}}{2q^{2}}$ in Eq. (\ref{eq:26}) and the nonflip contributions are the items without relation with this factor.

After finishing the integration of $\mathrm{cos}\theta$, we illustrate the dependence of these physical observables $A_{FB}^{l}(q^{2})$, $P_z^{F}(q^{2})$ and $P_z^{l}(q^{2})$ on the $q^{2}$ in Figs. \ref{AFB}-\ref{PZL}. From these figures, we can firstly see that these observables for the $b\rightarrow q$ and $c\rightarrow q$ transition process show very different characteristic. For example, the values of $A_{FB}^{l}(q^{2})$ for $c\rightarrow q$ transition are positive over the whole $q^{2}$ range, however the values are negative for the $b\rightarrow q$ process (See Fig. \ref{AFB}). Secondly, because $\tau$ have the larger mass than $\mu$ and $e$, the $\tau$ decaying mode (See green lines in these figures) is also different to the $\mu$- or $e$-mode. Thirdly, because the masses of the electron and muon are nearly equal to be zero, the longitudinal polarizations of these lepton are almost to be $-100\%$ and $100\%$ over the most $q^{2}$ range for $b\rightarrow q$ and $c\rightarrow q$ transition process (See Fig. \ref{PZL}), respectively. This changes only for $q^{2}$-values very close to the threshold $q^{2}=m_{l}^{2}$. However, for the $\tau$-mode, it has apparently different feature, and its absolute value gradually increase with $q^2$. Finally, by finishing the $q^{2}$-integration with $m_{l}^{2}\leq q^{2}\leq (m_{\mathcal{B}_{Q_{1}Q_{2}}}-m_{\mathcal{B}_{Q_{1}}^{*}})^{2}$, we can also obtain the values of $\langle A_{FB}^{l}\rangle$, $\langle P_z^{F}\rangle$ and $\langle P_z^{l}\rangle$ which are listed in the Tabs. \ref{result1} and \ref{result2}. We can see that the values of $\langle A_{FB}^{l}\rangle$ for $c\rightarrow q$ and $b\rightarrow q$ transition processes are in the range $0.21\sim0.31$ and $-0.56\sim-0.35$, respectively. The predicted values of $\langle P_z^{F}\rangle$ for $c\rightarrow q$ transition are apparently lower than those of the $b\rightarrow q$ transition. For the longitudinal polarization $\langle P_z^{l}\rangle$ of the lepton, their values of $\tau$-mode vary from $-0.69\sim-0.72$. However, the other values are almost to be $-1$ or $1$ because the masses of $e$ and $\mu$ are close to be zero. We hope these characteristics can provide useful information to study the weak decaying properties of the doubly heavy baryons in experiments.

\section{Conclusions}\label{sec5}
In the present work, we firstly analyze the form factors of $\mathcal{B}_{Q_{1}Q_{2}}(\frac{1}{2}^{+})\rightarrow\mathcal{B}_{Q_{1}}^{*}(\frac{3}{2}^{+})$ in the framework of three-point QCDSR. To eliminate the contaminations of $J^{P}=\frac{1}{2}^{-}$ state to mother baryon and $\frac{1}{2}^{\pm}$, $\frac{3}{2}^{-}$ states to daughter baryon, we establish 16 linear equations according to different dirac structures. By solving these equations, we successfully extract the form factors for this transition process. When performing the OPE in the QCD side, we include the perturbative part and vacuum condensate terms up to dimension 6. With the predicted form factors, we systematically analyze the semileptonic decays $\mathcal{B}_{Q_{1}Q_{2}}(\frac{1}{2}^{+})\rightarrow\mathcal{B}_{Q_{1}}^{*}(\frac{3}{2}^{+})$. Besides of the partial widths, ratios of $\Gamma_{L}/\Gamma_{T}$ and the branching fractions, some observables are also evaluated such as the forward-backward asymmetry paramter $A_{FB}^{l}$, the $P_z^{F}$ of the polarization vector for daughter baryons $\mathcal{B}_{Q_{1}}^{*}$ and the longitudinal polarization of the lepton $P_z^{l}$. The predictions in this work can deepen our understanding of the dynamics in the decay processes of doubly heavy baryons and provide useful information to explore the possibility of new physics in heavy baryonic decay channels.

\section{Acknowledgements}
This project is supported by National Natural Science Foundation under the Grant No. 12575083, and Natural Science Foundation of HeBei Province under the Grant No. A2024202008.
\clearpage
\begin{widetext}
\appendix
\section{Variations of the form factors with respect to the Borel parameters $M_{1}^{2}$ and $M_{2}^{2}$}\label{Sec:AppA}
In these following figures, the marked blue area denote the Borel platform used to extract the form factors.
\begin{figure*}[htbp]
	\centering
	\includegraphics[width=1.0\textwidth]{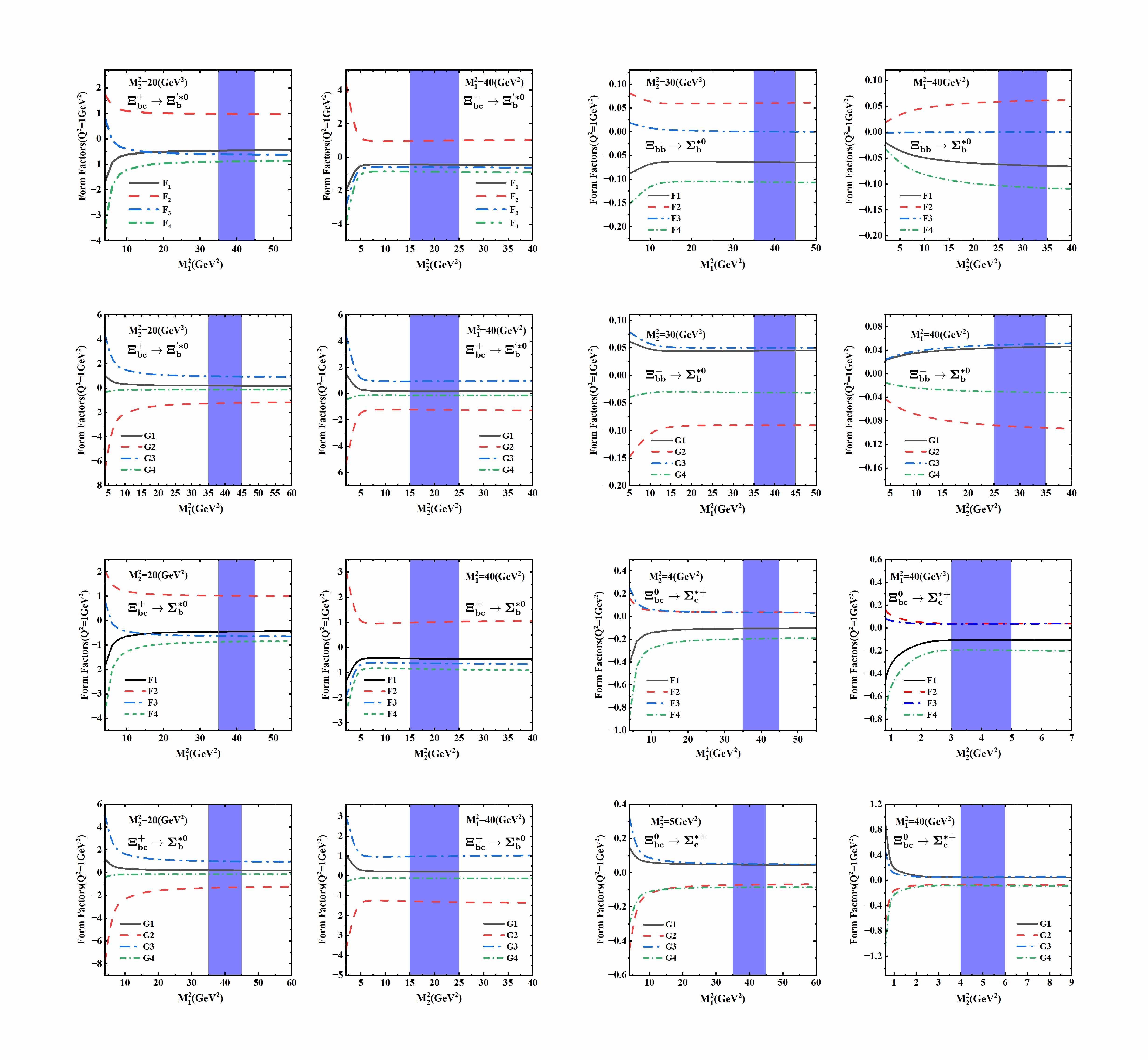}
	\caption{Variations of the form factors with respect to the Borel parameters for transition processes $\Xi_{bc}^{+}\rightarrow\Xi_{b}^{\prime*0}$, $\Xi_{bc}^{+}\rightarrow\Sigma_{b}^{*0}$, $\Xi_{bb}^{-}\rightarrow\Sigma_{b}^{*0}$ and $\Xi_{bc}^{0}\rightarrow\Sigma_{c}^{*+}$.}
\label{Fig15}
\end{figure*}
\begin{figure*}[htbp]
	\centering
	\includegraphics[width=1.0\textwidth]{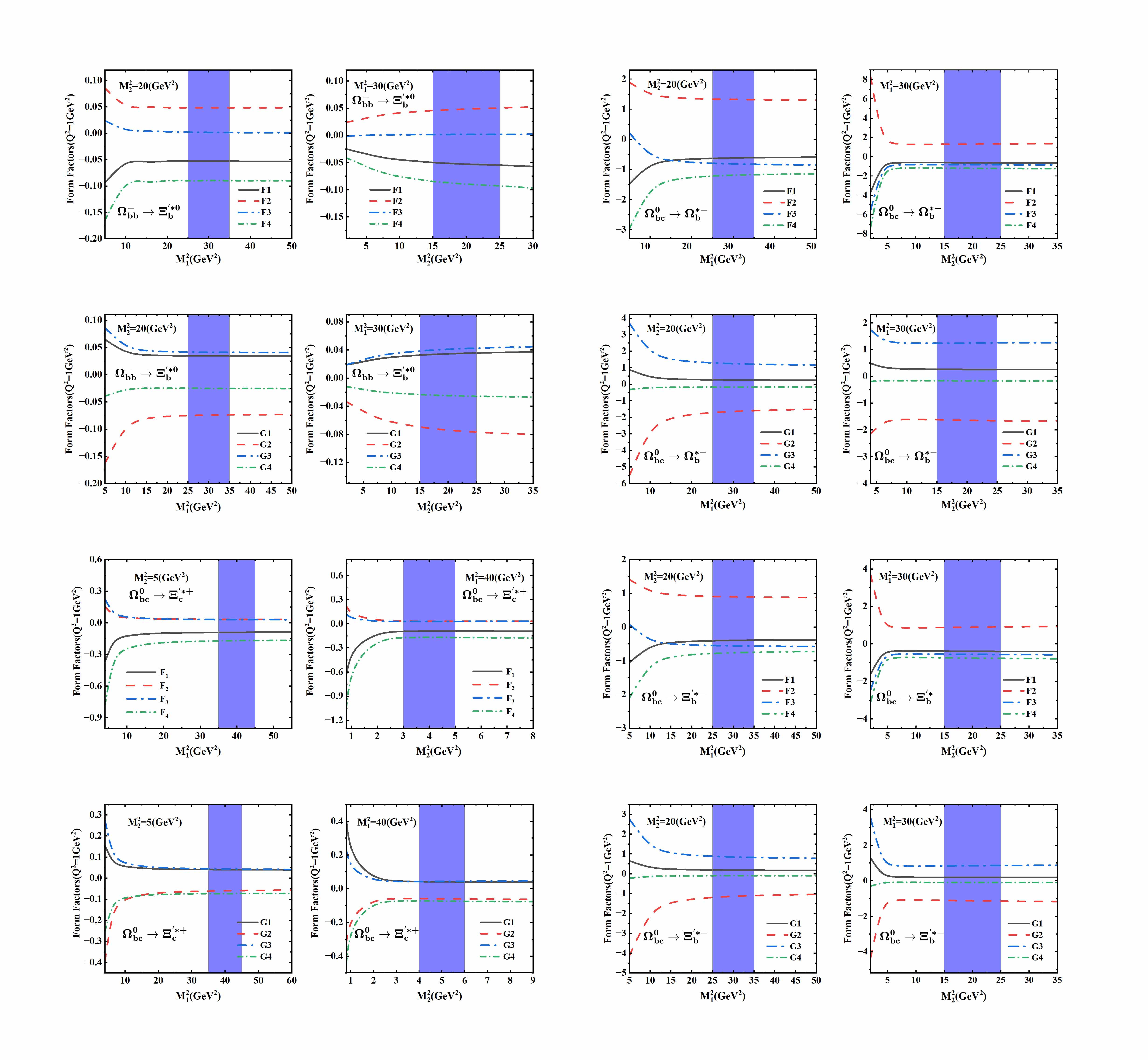}
	\caption{Variations of the form factors with respect to the Borel parameters for transition processes $\Omega_{bb}^{-}\rightarrow\Xi_{b}^{\prime*0}$, $\Omega_{bc}^{0}\rightarrow\Xi_{c}^{\prime*+}$, $\Omega_{bc}^{0}\rightarrow\Omega_{b}^{*-}$ and $\Omega_{bc}^{0}\rightarrow\Xi_{b}^{\prime*-}$.}
\label{Fig16}
\end{figure*}
\clearpage
\section{Variations of differential decay width with respect to $q^2$}\label{Sec:AppB}
\begin{figure*}[htbp] \centering
\begin{subfigure}{\includegraphics[width=0.8\textwidth]{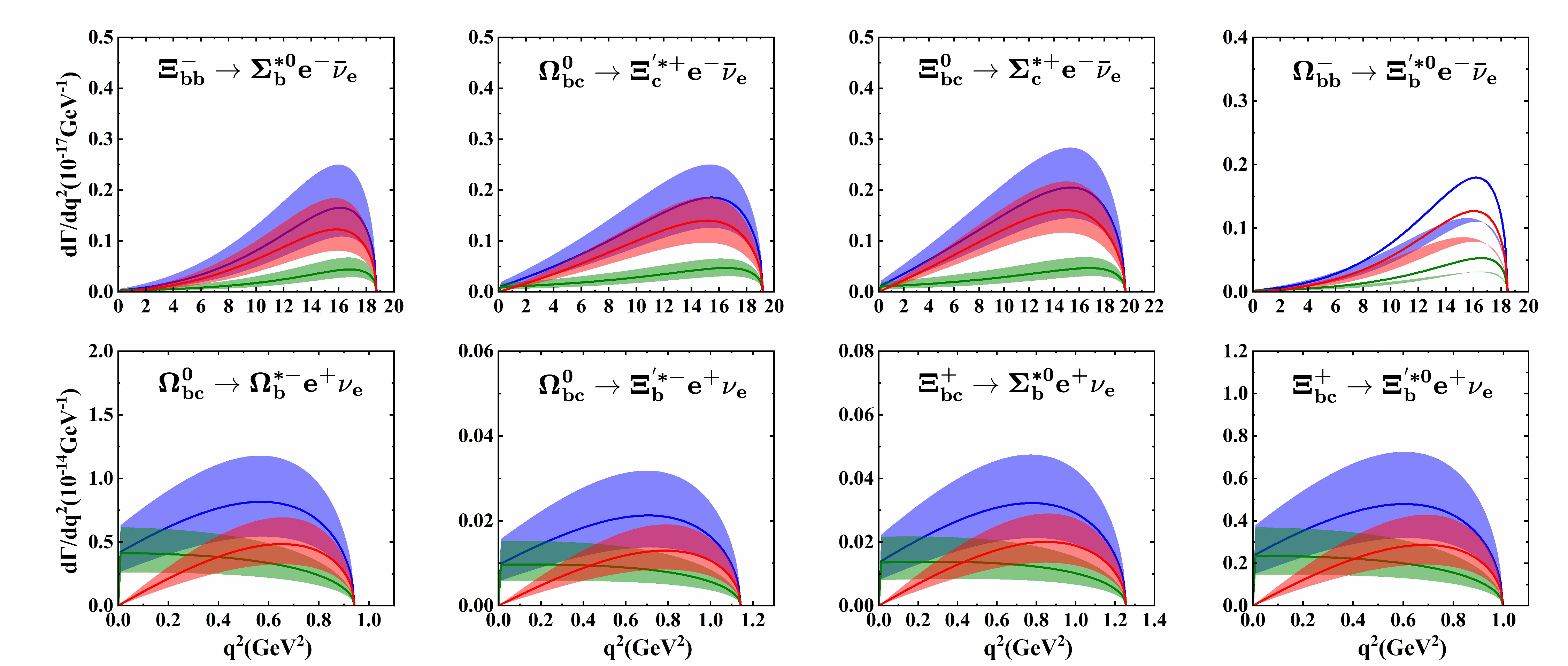}}
\caption{Variations of the differential decay width with respect to $q^2$ for the transition $\mathcal{B}_{Q_{1}Q_{2}}\rightarrow\mathcal{B}_{Q_{1}}^{*}$ with $e$-mode}
\label{decay1}
\end{subfigure}
\begin{subfigure}{\includegraphics[width=0.8\textwidth]{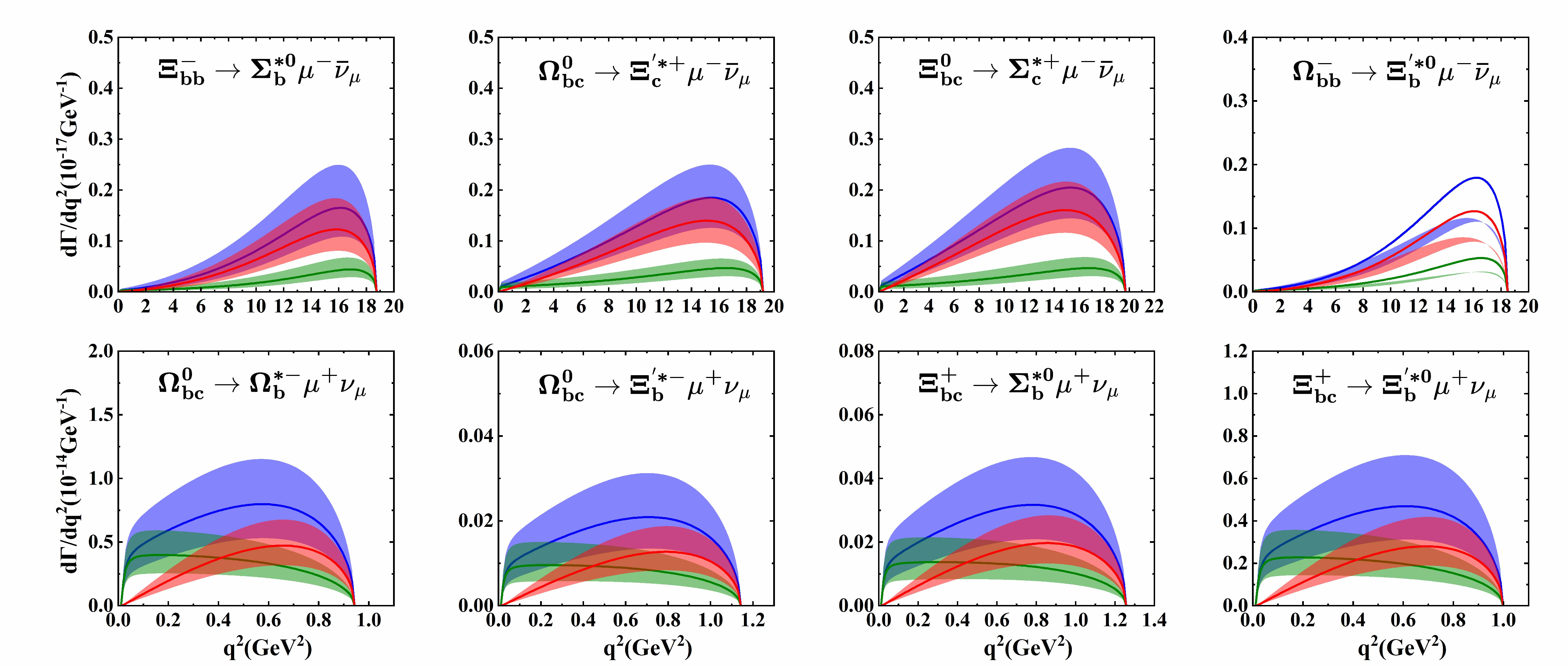}}
\caption{Variations of the differential decay width with respect to $q^2$ for the transition $\mathcal{B}_{Q_{1}Q_{2}}\rightarrow\mathcal{B}_{Q_{1}}^{*}$ with $\mu$-mode.}
\label{decay2}
\end{subfigure}
\begin{subfigure}{\includegraphics[width=0.8\textwidth]{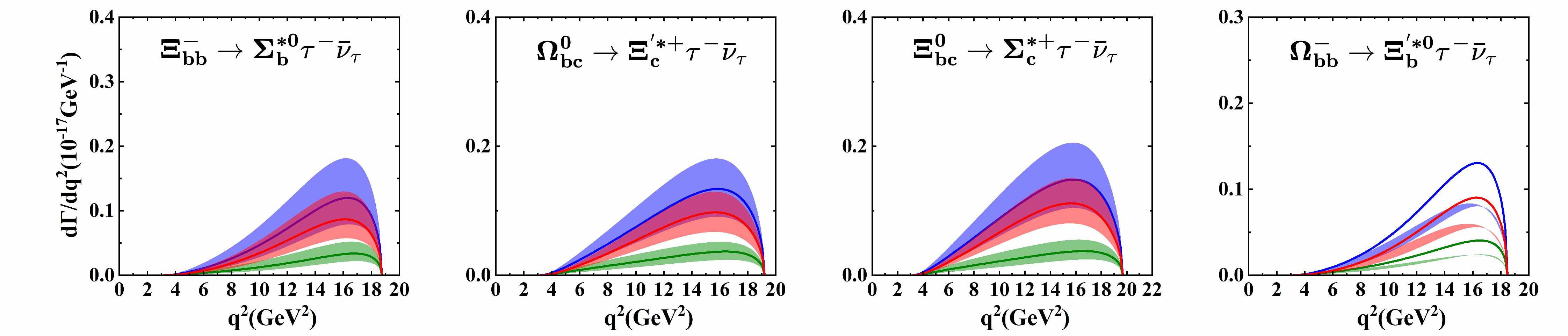}}
\caption{Variations of the differential decay width with respect to $q^2$ for the transition $\mathcal{B}_{Q_{1}Q_{2}}\rightarrow\mathcal{B}_{Q_{1}}^{*}$ with $\tau$-mode.}
\label{decay3}
\end{subfigure}
\end{figure*}
\clearpage
\section{Dependence of observable with respect to $q^2$}\label{Sec:AppC}
\begin{figure*}[htbp]
	\centering
	\includegraphics[width=15cm,height=5.4cm]{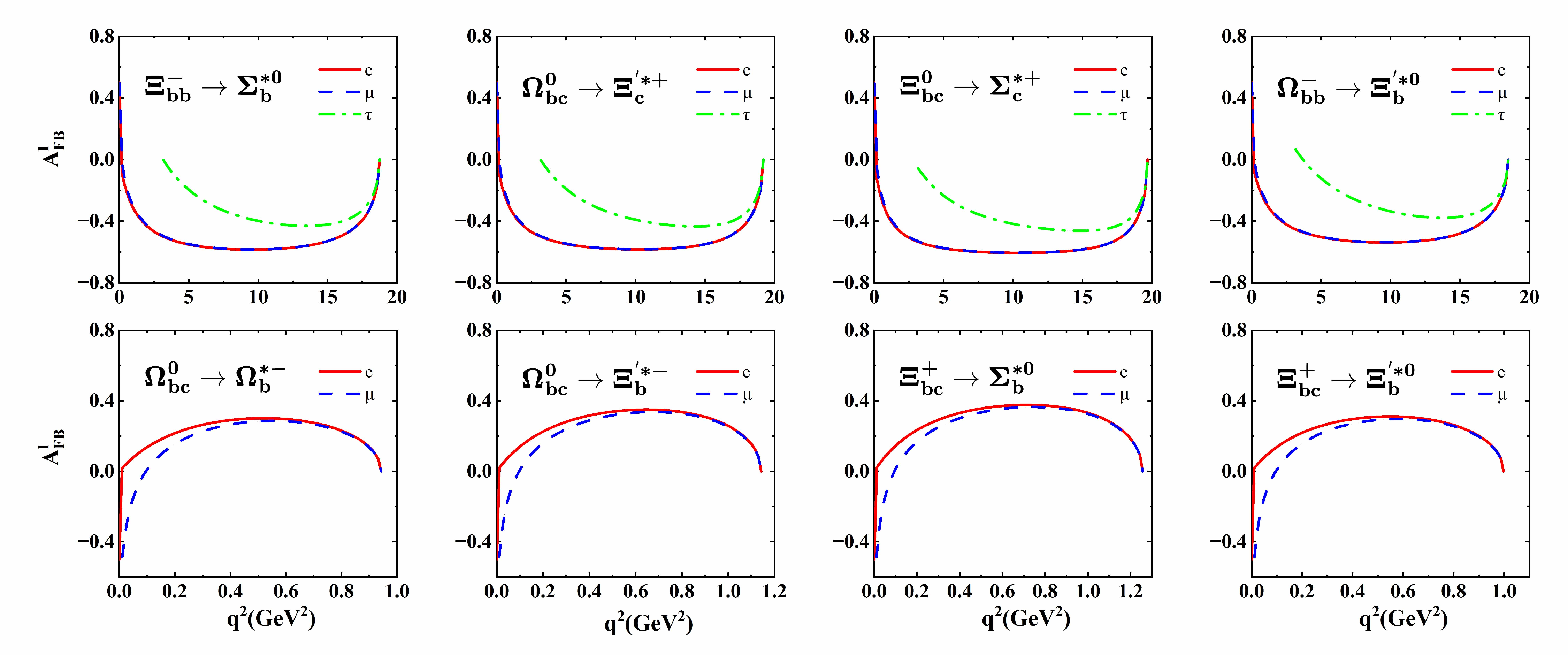}
	\caption{Dependence of forward-backward asymmetry $A_{FB}^{l}$ of the lepton on $q^{2}$.}
\label{AFB}
\end{figure*}
\begin{figure*}[htbp]
	\centering
	\includegraphics[width=15cm,height=5.4cm]{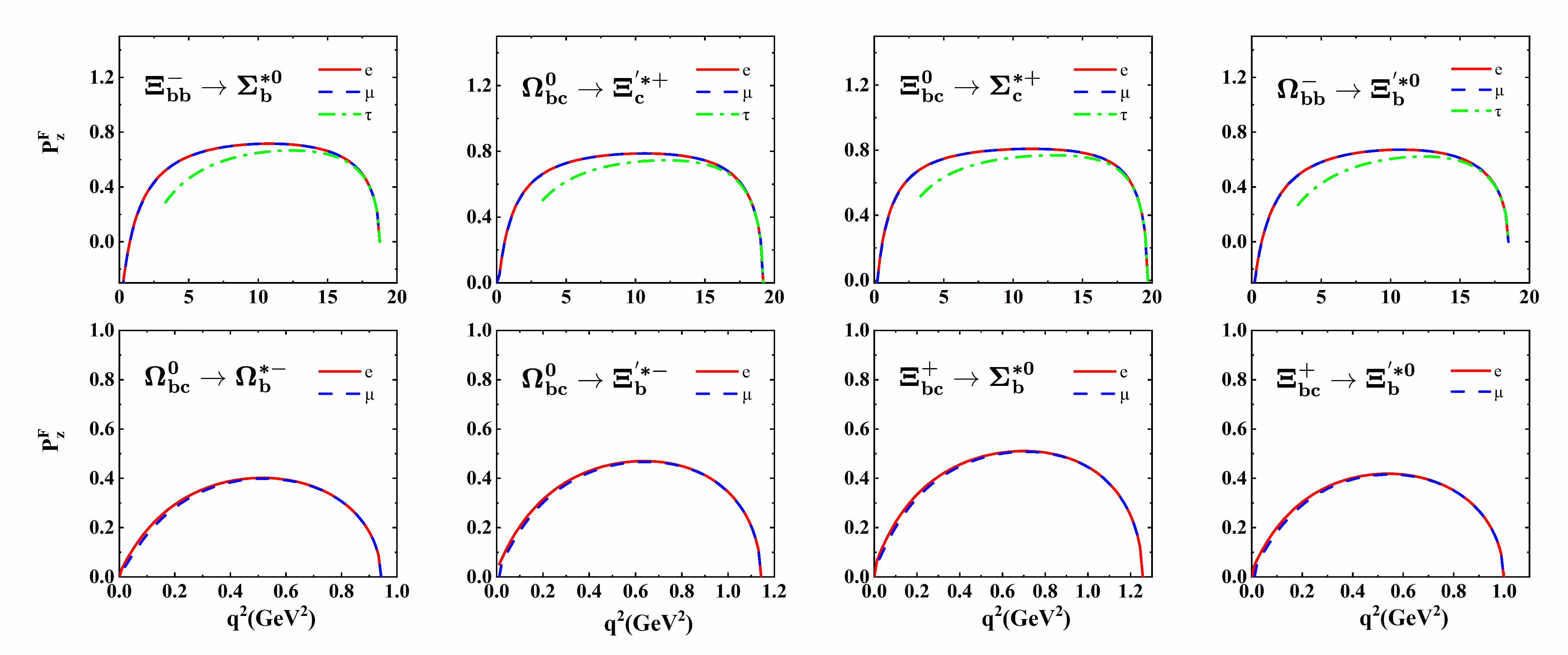}
	\caption{Same as Fig. 13 but for the component $P_z^{F}(\theta)$ of the polarization vector of daughter baryon.}
\label{PZF}
\end{figure*}
\begin{figure*}[htbp]
	\centering
	\includegraphics[width=15cm,height=5.4cm]{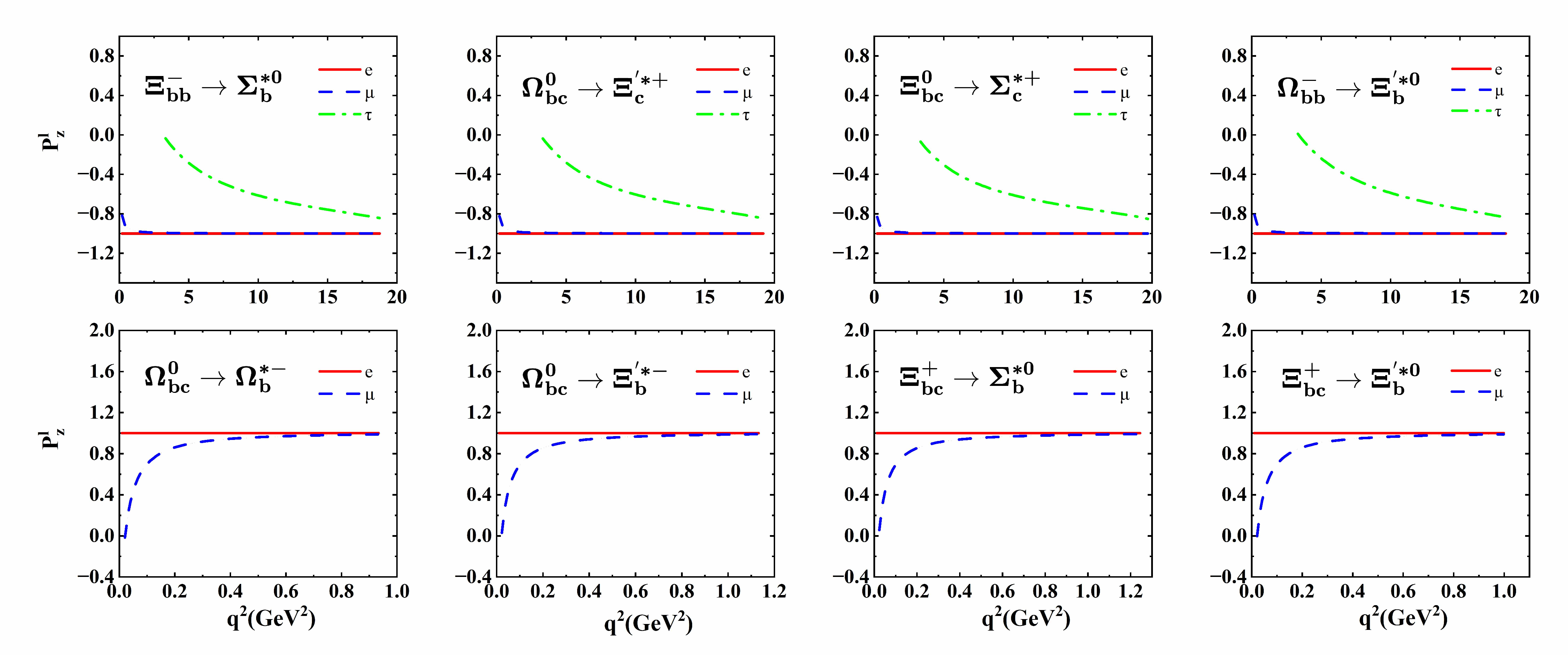}
	\caption{Same as Fig. 13 but for the longitudinal polarization $P_z^{l}$ of the lepton.}
\label{PZL}
\end{figure*}
\clearpage
\end{widetext}


\begin{thebibliography}{99}
\bibitem{Ebert:2002ig}
D.~Ebert, R.~N.~Faustov, V.~O.~Galkin and A.~P.~Martynenko,
\href{https://doi:10.1103/PhysRevD.66.014008}{Phys. Rev. D \textbf{66}, 014008 (2002)}.

\bibitem{Roberts:2007ni}
W.~Roberts and M.~Pervin,
\href{https://doi:10.1142/S0217751X08041219}{Int. J. Mod. Phys. A \textbf{23}, 2817-2860 (2008)}.

\bibitem{Yu:2022lel}
G.~L.~Yu, Z.~Y.~Li, Z.~G.~Wang, J.~Lu and M.~Yan,
\href{https://doi:10.1140/epja/s10050-023-01044-1}{Eur. Phys. J. A \textbf{59}, 126 (2023)}.

\bibitem{Li:2022ywz}
Z.~Y.~Li, G.~L.~Yu, Z.~G.~Wang, J.~Z.~Gu and H.~T.~Shen,
\href{https://doi:10.1142/S0217751X23500951}{Int. J. Mod. Phys. A \textbf{38}, 2350095 (2023)}.

\bibitem{Yu:2022ymb}
G.~L.~Yu, Z.~Y.~Li, Z.~G.~Wang, J.~Lu and M.~Yan,
\href{https://doi:10.1016/j.nuclphysb.2023.116183}{Nucl. Phys. B \textbf{990}, 116183 (2023)}.

\bibitem{Li:2022oth}
Z.~Y.~Li, G.~L.~Yu, Z.~G.~Wang, J.~Z.~Gu and H.~T.~Shen,
\href{https://doi:10.1142/S0217732323500529}{Mod. Phys. Lett. A \textbf{38}, 2350052 (2023)}.

\bibitem{Li:2022xtj}
Z.~Y.~Li, G.~L.~Yu, Z.~G.~Wang, J.~Z.~Gu, J.~Lu and H.~T.~Shen,
\href{https://doi:10.1088/1674-1137/acd365}{Chin. Phys. C \textbf{47}, 073105 (2023)}.

\bibitem{Karliner:2014gca}
M.~Karliner and J.~L.~Rosner,
\href{https://doi:10.1103/PhysRevD.90.094007}{Phys. Rev. D \textbf{90}, 094007 (2014)}.

\bibitem{Valcarce:2008dr}
A.~Valcarce, H.~Garcilazo and J.~Vijande,
\href{https://doi:10.1140/epja/i2008-10616-4}{Eur. Phys. J. A \textbf{37}, 217 (2008)}.

\bibitem{Bagan:1992za}
E.~Bagan, M.~Chabab and S.~Narison,
\href{https://doi:10.1016/0370-2693(93)90090-5}{Phys. Lett. B \textbf{306}, 350 (1993)}.

\bibitem{Wang:2010hs}
Z.~G.~Wang,
\href{https://doi:10.1140/epja/i2010-11004-3}{Eur. Phys. J. A \textbf{45}, 267 (2010)}.

\bibitem{Wang:2010vn}
Z.~G.~Wang,
\href{https://doi:10.1140/epjc/s10052-010-1357-8}{Eur. Phys. J. C \textbf{68}, 459 (2010)}.


\bibitem{Kiselev:2001fw}
V.~V.~Kiselev and A.~K.~Likhoded,
\href{https://doi:10.1070/PU2002v045n05ABEH000958}{Phys. Usp. \textbf{45}, 455 (2002)}.

\bibitem{Zhang:2008rt}
J.~R.~Zhang and M.~Q.~Huang,
\href{https://doi:10.1103/PhysRevD.78.094007}{Phys. Rev. D \textbf{78}, 094007 (2008)}.

\bibitem{Aliev:2012ru}
T.~M.~Aliev, K.~Azizi and M.~Savci,
\href{https://doi:10.1016/j.nuclphysa.2012.09.009}{Nucl. Phys. A \textbf{895}, 59 (2012)}.

\bibitem{He:2004px}
D.~H.~He, K.~Qian, Y.~B.~Ding, X.~Q.~Li and P.~N.~Shen,
\href{https://doi:10.1103/PhysRevD.70.094004}{Phys. Rev. D \textbf{70}, 094004 (2004)}.

\bibitem{Cheng:2026mlv}
H.~Y.~Cheng and C.~W.~Liu,
\href{https://doi.org/10.48550/arXiv.2604.10939}{arXiv:2604.10939 [hep-ph]}.

\bibitem{Lewis:2001iz}
R.~Lewis, N.~Mathur and R.~M.~Woloshyn,
\href{https://doi:10.1103/PhysRevD.64.094509}{Phys. Rev. D \textbf{64}, 094509 (2001)}.

\bibitem{Flynn:2003vz}
J.~M.~Flynn \textit{et al.} [UKQCD],
\href{https://doi:10.1088/1126-6708/2003/07/066}{JHEP \textbf{07}, 066 (2003)}.

\bibitem{Liu:2009jc}
L.~Liu, H.~W.~Lin, K.~Orginos and A.~Walker-Loud,
\href{https://doi:10.1103/PhysRevD.81.094505}{Phys. Rev. D \textbf{81}, 094505 (2010)}.

\bibitem{Meinel:2021mdj}
S.~Meinel and G.~Rendon,
\href{https://doi:10.1103/PhysRevD.105.054511}{Phys. Rev. D \textbf{105}, 054511 (2022)}.

\bibitem{352117}
R.~Aaij \textit{et al.} [LHCb],
\href{https://doi.org/10.1103/PhysRevLett.119.112001}{Phys. Rev. Lett. \textbf{119}, 112001 (2017)}.

\bibitem{352118}
R.~Aaij \textit{et al.} [LHCb],
\href{https://doi.org/10.1103/PhysRevLett.121.162002}{Phys. Rev. Lett. \textbf{121}, 162002 (2018)}.

\bibitem{352119}
R.~Aaij \textit{et al.} [LHCb],
\href{https://doi.org/10.1007/JHEP02(2020)049}{JHEP \textbf{02}, 049 (2020)}.

\bibitem{LHCb:2020iko}
R.~Aaij \textit{et al.} [LHCb],
\href{https://doi:10.1007/JHEP11(2020)095}{JHEP \textbf{11}, 095 (2020)}.

\bibitem{LHCb:2021xba}
R.~Aaij \textit{et al.} [LHCb],
\href{https://doi:10.1088/1674-1137/ac0c70}{Chin. Phys. C \textbf{45}, 093002 (2021)}.

\bibitem{Faustov:2018ahb}
R.~N.~Faustov and V.~O.~Galkin,
\href{https://doi:10.1103/PhysRevD.98.093006}{Phys. Rev. D \textbf{98}, 093006 (2018)}.

\bibitem{Geng:2020ofy}
C.~Q.~Geng, C.~W.~Liu and T.~H.~Tsai,
\href{https://doi:10.1103/PhysRevD.102.034033}{Phys. Rev. D \textbf{102}, 034033 (2020)}.

\bibitem{Albertus:2004wj}
C.~Albertus, E.~Hernandez and J.~Nieves,
\href{https://doi:10.1103/PhysRevD.71.014012}{Phys. Rev. D \textbf{71}, 014012 (2005)}.

\bibitem{Ebert:2006rp}
D.~Ebert, R.~N.~Faustov and V.~O.~Galkin,
\href{https://doi:10.1103/PhysRevD.73.094002}{Phys. Rev. D \textbf{73}, 094002 (2006)}.

\bibitem{Cheng:1996cs}
H.~Y.~Cheng,
\href{https://doi:10.1103/PhysRevD.56.2799}{Phys. Rev. D \textbf{56}, 2799 (1997)}.

\bibitem{Ivanov:1997hi}
M.~A.~Ivanov, J.~G.~Korner, V.~E.~Lyubovitskij and A.~G.~Rusetsky,
\href{https://doi:10.1142/S0217732398000231}{Mod. Phys. Lett. A \textbf{13}, 181 (1998)}.

\bibitem{Ivanov:1997ra}
M.~A.~Ivanov, J.~G.~Korner, V.~E.~Lyubovitskij and A.~G.~Rusetsky,
\href{https://doi:10.1103/PhysRevD.57.5632}{Phys. Rev. D \textbf{57}, 5632 (1998)}.

\bibitem{Faessler:2009xn}
A.~Faessler, T.~Gutsche, M.~A.~Ivanov, J.~G.~Korner and V.~E.~Lyubovitskij,
\href{https://doi:10.1103/PhysRevD.80.034025}{Phys. Rev. D \textbf{80}, 034025 (2009)}.

\bibitem{Gutsche:2018utw}
T.~Gutsche, M.~A.~Ivanov, J.~G.~K\"orner and V.~E.~Lyubovitskij,
\href{https://doi:10.1103/PhysRevD.98.074011}{Phys. Rev. D \textbf{98}, 074011 (2018)}.

\bibitem{Gutsche:2019iac}
T.~Gutsche, M.~A.~Ivanov, J.~G.~K{\"o}rner, V.~E.~Lyubovitskij and Z.~Tyulemissov,
\href{https://doi:10.1103/PhysRevD.100.114037}{Phys. Rev. D \textbf{100}, 114037 (2019)}.

\bibitem{Wang:2017azm}
W.~Wang, Z.~P.~Xing and J.~Xu,
\href{https://doi:10.1140/epjc/s10052-017-5363-y}{Eur. Phys. J. C \textbf{77}, 800 (2017)}.

\bibitem{Albertus:2013wja}
C.~Albertus, E.~Hernandez and J.~Nieves,
\href{https://doi:10.1007/s00601-013-0739-5}{Few Body Syst. \textbf{55}, 767 (2014)}.

\bibitem{Zhao:2018mrg}
Z.~X.~Zhao,
\href{https://doi:10.1140/epjc/s10052-018-6213-2}{Eur. Phys. J. C \textbf{78}, 756 (2018)}.

\bibitem{Zhao:2018zcb}
Z.~X.~Zhao,
\href{https://doi:10.1088/1674-1137/42/9/093101}{Chin. Phys. C \textbf{42}, 093101 (2018)}.

\bibitem{Chua:2018lfa}
C.~K.~Chua,
\href{https://doi:10.1103/PhysRevD.99.014023}{Phys. Rev. D \textbf{99}, 014023 (2019)}.

\bibitem{Chua:2019yqh}
C.~K.~Chua,
\href{https://doi:10.1103/PhysRevD.100.034025}{Phys. Rev. D \textbf{100}, 034025 (2019)}.

\bibitem{Ke:2019smy}
H.~W.~Ke, N.~Hao and X.~Q.~Li,
\href{https://doi:10.1140/epjc/s10052-019-7048-1}{Eur. Phys. J. C \textbf{79} (2019), 540}.

\bibitem{Zhu:2018jet}
J.~Zhu, Z.~T.~Wei and H.~W.~Ke,
\href{https://doi:10.1103/PhysRevD.99.054020}{Phys. Rev. D \textbf{99}, 054020 (2019)}.

\bibitem{Li:2021qod}
Y.~S.~Li, X.~Liu and F.~S.~Yu,
\href{https://}{Phys. Rev. D \textbf{104}, 013005 (2021)}.

\bibitem{Hu:2020mxk}
X.~H.~Hu, R.~H.~Li and Z.~P.~Xing,
\href{https://doi:10.1140/epjc/s10052-020-7851-8}{Eur. Phys. J. C \textbf{80}, 320 (2020)}.

\bibitem{Li:2021kfb}
Y.~S.~Li and X.~Liu,
\href{https://doi:10.1103/PhysRevD.105.013003}{Phys. Rev. D \textbf{105}, 013003 (2022)}.

\bibitem{Lu:2023rmq}
F.~Lu, H.~W.~Ke, X.~H.~Liu and Y.~L.~Shi,
\href{https://doi:10.1140/epjc/s10052-023-11572-1}{Eur. Phys. J. C \textbf{83}, 412 (2023)}.

\bibitem{Liu:2022mxv}
H.~Liu, Z.~P.~Xing and C.~Yang,
\href{https://doi:10.1140/epjc/s10052-023-11263-x}{Eur. Phys. J. C \textbf{83}, 123 (2023)}.

\bibitem{Aliev:2010uy}
T.~M.~Aliev, K.~Azizi and M.~Savci,
\href{https://doi:10.1103/PhysRevD.81.056006}{Phys. Rev. D \textbf{81}, 056006 (2010)}.

\bibitem{Aliev:2023tpk}
T.~M.~Aliev, S.~Bilmis and M.~Savci,
\href{https://doi:10.1016/j.physletb.2023.138287}{Phys. Lett. B \textbf{847}, 138287 (2023)}.

\bibitem{Azizi:2011mw}
K.~Azizi, Y.~Sarac and H.~Sundu,
\href{https://doi:10.1140/epja/i2012-12002-1}{Eur. Phys. J. A \textbf{48}, 2 (2012)}.

\bibitem{Wang:2008sm}
Y.~m.~Wang, Y.~Li and C.~D.~Lu,
\href{https://doi:10.1140/epjc/s10052-008-0846-5}{Eur. Phys. J. C \textbf{59}, 861 (2009)}.

\bibitem{Wang:2015ndk}
Y.~M.~Wang and Y.~L.~Shen,
\href{https://doi:10.1007/JHEP02(2016)179}{JHEP \textbf{02}, 179 (2016)}.

\bibitem{Khodjamirian:2011jp}
A.~Khodjamirian, C.~Klein, T.~Mannel and Y.~M.~Wang,
\href{https://doi:10.1007/JHEP09(2011)106}{JHEP \textbf{09}, 106 (2011)}.

\bibitem{Zhao:2020mod}
Z.~X.~Zhao, R.~H.~Li, Y.~L.~Shen, Y.~J.~Shi and Y.~S.~Yang,
\href{https://doi:10.1140/epjc/s10052-020-08767-1}{Eur. Phys. J. C \textbf{80}, 1181 (2020)}.

\bibitem{Shi:2019hbf}
Y.~J.~Shi, W.~Wang and Z.~X.~Zhao,
\href{https://doi:10.1140/epjc/s10052-020-8096-2}{Eur. Phys. J. C \textbf{80}, 568 (2020)}.

\bibitem{Khajouei:2024frw}
L.~Khajouei and K.~Azizi,
\href{https://doi:10.1103/PhysRevD.111.074018}{Phys. Rev. D \textbf{111}, 074018 (2025)}.

\bibitem{Miao:2022bga}
Y.~Miao, H.~Deng, K.~S.~Huang, J.~Gao and Y.~L.~Shen,
\href{https://doi:10.1088/1674-1137/ac8652}{Chin. Phys. C \textbf{46}, 113107 (2022)}.

\bibitem{Xing:2021enr}
Z.~P.~Xing and Z.~X.~Zhao,
\href{https://doi:10.1140/epjc/s10052-021-09902-2}{Eur. Phys. J. C \textbf{81}, 1111 (2021)}.

\bibitem{Zhao:2021sje}
Z.~X.~Zhao, X.~Y.~Sun, F.~W.~Zhang, Y.~P.~Xing and Y.~T.~Yang,
\href{https://doi:10.1103/PhysRevD.108.116008}{Phys. Rev. D \textbf{108}, 116008 (2023)}.

\bibitem{Lu:2026qkk}
J.~Lu, G.~L.~Yu, D.~Y.~Chen, Z.~G.~Wang and B.~Wu,
\href{https://doi:10.1140/epjc/s10052-026-15871-1}{Eur. Phys. J. C \textbf{86}, 685 (2026)}.

\bibitem{Zhang:2023nxl}
S.~Q.~Zhang and C.~F.~Qiao,
\href{https://doi:10.1103/PhysRevD.108.074017}{Phys. Rev. D \textbf{108}, 074017 (2023)}.

\bibitem{Neishabouri:2024gbc}
Z.~Neishabouri, K.~Azizi and H.~R.~Moshfegh,
\href{https://doi:10.1103/PhysRevD.110.014010}{Phys. Rev. D \textbf{110}, 014010 (2024)}.

\bibitem{Tousi:2024usi}
M.~S.~Tousi, K.~Azizi and H.~R.~Moshfegh,
\href{https://doi:10.1103/PhysRevD.110.114001}{Phys. Rev. D \textbf{110}, 114001 (2024)}.

\bibitem{Yu:2026tbk}
G.~L.~Yu, Z.~G.~Wang, J.~Lu, B.~Wu, P.~Yang and Z.~Zhou,
\href{https://doi:10.1103/ghm9-2lmd}{Phys. Rev. D \textbf{113}, 7 (2026)}.

\bibitem{Shifman:1978bx}
M.~A.~Shifman, A.~I.~Vainshtein and V.~I.~Zakharov,
\href{https://doi:10.1016/0550-3213(79)90022-1}{Nucl. Phys. B \textbf{147}, 385 (1979)}.

\bibitem{Shifman:1978by}
M.~A.~Shifman, A.~I.~Vainshtein and V.~I.~Zakharov,
\href{https://doi:10.1016/0550-3213(79)90023-3}{Nucl. Phys. B \textbf{147}, 448 (1979)}.

\bibitem{Colangelo:2000dp}
P.~Colangelo and A.~Khodjamirian,
\href{https://doi:10.1142/9789812810458{\_}0033}{arXiv:hep-ph/0010175}.

\bibitem{Wang:2025sic}
Z.~G.~Wang,
\href{https://doi:10.15302/frontphys.2026.016300}{Front. Phys. (Beijing) \textbf{21}, 016300 (2026)}.

\bibitem{Wang:2022ufh}
Z.~G.~Wang and Q.~Xin,
\href{https://doi:10.1142/S0217751X22500749}{Int. J. Mod. Phys. A \textbf{37}, 2250074 (2022)}.

\bibitem{Wang:2018lhz}
Z.~G.~Wang,
\href{https://doi:10.1140/epjc/s10052-018-6300-4}{Eur. Phys. J. C \textbf{78}, 826 (2018)}.

\bibitem{Wang:2017vtv}
Z.~G.~Wang,
\href{https://doi:10.1016/j.nuclphysb.2017.11.014}{Nucl. Phys. B \textbf{926}, 467 (2018)}.

\bibitem{Zhang:2025qmg}
S.~Q.~Zhang and C.~F.~Qiao,
\href{https://doi:10.1007/s43673-026-00192-y}{AAPPS Bull. \textbf{36}, 12 (2026)}.

\bibitem{Lu:2025zaf}
J.~Lu, D.~Y.~Chen, G.~L.~Yu, Z.~G.~Wang and Z.~Zhou,
\href{https://doi:10.1140/epjc/s10052-025-14754-1}{Eur. Phys. J. C \textbf{85}, 1061 (2025)}.

\bibitem{Lu:2025gol}
J.~Lu, G.~L.~Yu, D.~Y.~Chen, Z.~G.~Wang and B.~Wu,
\href{https://doi:10.1140/epjc/s10052-025-15110-z}{Eur. Phys. J. C \textbf{85}, 1382 (2025)}.

\bibitem{Cutkosky}
R.E. Cutkosky, J. Math. Phys. 1, 429 (1960).

\bibitem{ParticleDataGroup:2024cfk}
S.~Navas \textit{et al.} [Particle Data Group],
\href{https://doi:10.1103/PhysRevD.110.030001}{Phys. Rev. D \textbf{110}, 030001 (2024)}.

\bibitem{Pascual}
P.Pascual,R.Tarrach, QCD:Renormalization for the Practitioner, 194 (1984).

\bibitem{Reinders}
L.J. Reinders, H. Rubinstein, S. Yazaki, Phys. Rep. 127, 1 (1985).

\bibitem{Boyd:1994tt}
C.~G.~Boyd, B.~Grinstein and R.~F.~Lebed,
\href{https://doi:10.1103/PhysRevLett.74.4603}{Phys. Rev. Lett. \textbf{74}, 4603 (1995)}.

\bibitem{Narison:2010cg}
S.~Narison,
\href{https://doi:10.1016/j.physletb.2011.09.116}{Phys. Lett. B \textbf{693}, 559 (2010)}.

\bibitem{Narison:2011xe}
S.~Narison,
\href{https://doi:10.1016/j.physletb.2011.11.058}{Phys. Lett. B \textbf{706}, 412 (2012)}.

\bibitem{Narison:2011rn}
S.~Narison,
\href{https://doi:10.1016/j.physletb.2011.12.047}{Phys. Lett. B \textbf{707}, 259 (2012)}.

\bibitem{Hu:2022xzu}
X.~H.~Hu and Y.~J.~Shi,
\href{https://doi:10.1103/PhysRevD.107.036007}{Phys. Rev. D \textbf{107}, 036007 (2023)}.

\bibitem{Kadeer:2005aq}
A.~Kadeer, J.~G.~Korner and U.~Moosbrugger,
\href{https://doi:10.1140/epjc/s10052-008-0801-5}{Eur. Phys. J. C \textbf{59}, 27 (2009)}.
\end{thebibliography}
\end{document}